\documentclass[]{aa}  

\usepackage{graphicx}
\usepackage{txfonts}
\usepackage{hyperref}
\usepackage{mhchem}
\usepackage{lastpage}
\usepackage{color}

\newcommand{\Rp}{R_{\rm p}}
\newcommand{\Rs}{R_{\rm s}}
\newcommand{\Mp}{M_{\rm p}}
\newcommand{\Mearth}{M_{\oplus}}
\newcommand{\MJ}{{\,{\rm M_{\rm J}}}}
\newcommand{\RJ}{{\,{\rm R_{\rm J}}}}
\newcommand{\Rsun}{{\,R_{\odot}}}
\newcommand{\cs}{c_{\rm s}}

\begin{document}

   \title{Analytical modeling of helium absorption signals of isothermal atmospheric escape }

\author{H. Mitani
\inst{1,2} \thanks{E-mail: hiroto.mitani@uni-due.de}
\and
R. Kuiper
\inst{1}
}
\institute{
Faculty of Physics, University of Duisburg-Essen, Lotharstra{\ss}e 1, D-47057 Duisburg, Germany
\and
Department of Physics, School of Science, The University of Tokyo, 7-3-1 Hongo, Bunkyo, Tokyo 113-0033, Japan
}

  \abstract
   { Atmospheric escape driven by extreme ultraviolet (EUV) radiation is a critical process shaping the evolution of close-in exoplanets. 
Recent observations have detected helium triplet absorption in numerous (>20) close-in exoplanets, highlighting the importance of understanding upper atmospheric thermo-chemical structure. While super-solar metallicity has been observed in the atmospheres of some close-in exoplanets, the impact of metal species on both atmospheric escape dynamics and observed absorption features remains poorly understood.  
In this study, we derive a simplified yet accurate formula for the equivalent width of helium absorption in the limit of an isothermal temperature for the upper atmosphere. Our results demonstrate that planets with lower temperature (metal-rich atmosphere) exhibit lower mass-loss rate although the equivalent width of helium triplet absorption remains largely independent of atmospheric temperature (metallicity) because the low temperatures in these atmospheres enhance the fraction of helium in its triplet state. Additionally, we present a hydrodynamic model based on radiation-hydrodynamics simulations that incorporates the effects of metal cooling. Our analytical model can predict the helium triplet equivalent width of the atmosphere of simulations.
The analytical model provides a comprehensive framework for understanding how metal cooling in the upper atmosphere influences the thermo-chemical structure and observable helium features of close-in exoplanetary atmospheres, offering valuable insights for interpreting current and future observational data.
   }
   
\titlerunning{Analytical model of Helium absorption}


   \maketitle

\section{Introduction}
Atmospheric escape driven by intense ultraviolet radiation from host stars is a critical process for understanding the evolution of hot close-in exoplanets \citep{Garcia_2007,Owen_2017,Owen_2019,Mazeh_2016,Fulton_2017}. Observations of escaping atmospheres have been made through Ly$\alpha$ absorption \citep{Vidal-Madjar_2003,Ehrenreich_2015,Bourrier_2018,Rockliffe_2021}, and constructing theoretical models of escaping atmospheres is essential for interpreting the observed absorption signals to infer atmospheric structure.
Detections of helium triplet absorption at $10830$\,\AA\, have provided a powerful tool for studying escaping atmospheres \citep{Oklopcic_2018,Allart_2018,Spake_2018,Zhang_2023c,Alam_2024}. Helium triplet absorption is observable with ground-based telescopes, and the growing number of exoplanets exhibiting this feature highlights the need for developing a general theoretical model applicable to a wide range of systems to understand the physics, which determine the atmospheric structure.

However, there are puzzling non-detections of helium absorption in some close-in exoplanets \citep{Zhang_2022,Bennett_2023,Allart_2023,Orell-Miquel_2024}. Several scenarios have been proposed to explain these non-detections. For instance, the absence of a primordial hydrogen/helium-dominated atmosphere could account for the lack of detectable helium absorption of certain planets \citep{Zhang_2022}. Additionally, stellar wind interactions play a significant role in shaping planetary outflows and influencing observed signals \citep{Mitani_2022,McCann_2019,Carolan_2020}. The geometry and extent of the escaping atmosphere can be altered by the interplay between the stellar wind and the planetary outflow. In particular, strong ram pressure from the stellar wind can confine the escaping atmosphere, thereby reducing observable absorption signals \citep{Mitani_2022}.

While numerous studies have focused on understanding observed helium absorptions in specific systems \citep[e.g.][]{Kirk_2022,Zhang_2023c}, there is a pressing need to develop a general theoretical framework that explains why certain planets exhibit helium non-detection inconsistent with simple hydrogen/helium-dominated atmosphere models. Recent theoretical studies constructed a general model for Ly$\alpha$/helium transit absorption under an energy-limited assumption \citep{Owen_2023,Schreyer_2024,Ballabio_2025}.  

Recent detections of metal line absorptions in the upper atmospheres of close-in exoplanets \citep{Vidal-Madjar_2013,Sing_2019,Ben-Jaffel_2022,Boehm_2025} suggest that metal cooling could significantly impact atmospheric escape and associated observational signals. However, many existing models of upper atmospheres primarily consider hydrogen/helium-dominated compositions. Recent several studies have started to include metal and molecular heating and cooling processes in models of close-in giants and sub-Neptunes \citep[e.g.,][]{Koskinen_2014,Koskinen_2022,Linssen_2022,Linssen_2024,Munoz_2023,Kubyshkina_2024,Yoshida_2025}. Nevertheless, the overall impact of metals on the structure and escape of close-in planetary upper atmospheres is still not fully understood.

In this paper, we focus on close-in planets where intense stellar radiation drives strong atmospheric escape. We developed an analytic model for hydrogen-dominated atmospheres incorporating metal cooling to explore how such systems align with radiation-hydrodynamics simulations. As one step to understand the observed helium absorption signal, we study the parameter dependence of the helium absorption.

\section{Helium triplet absorption}
\label{sec:He_analytic}
Helium triplet absorption at $10830$\,\AA\ is widely used to detect atmospheric escape.
Understanding absorption in simple isothermal atmospheres before simulations that take into account the detailed planetary atmospheric structure will provide an important foundation for understanding observations.

In this section, we derive analytic formulae of the equivalent width of helium absorption for an isothermal atmosphere.

\subsection{Transmission of helium absorption}
To calculate the equivalent width of absorption, we first calculate the transmission of the helium absorption.
The transmission $T_{\lambda}$ can be defined using the optical depth $\tau_{\lambda}(b)$ as
\begin{equation}
    T_{\lambda} = \frac{1}{\pi(\Rs^2-\Rp^2)} \int_{\Rp}^{\Rs} 2\pi b e^{-\tau_{\lambda}(b)} db
    \label{eq:transmission}
\end{equation}
where $R_s,\Rp$ are the stellar radius and the planetary radius.
The optical depth with the impact parameter $b$ is given by
\begin{equation}
    \tau_{\lambda}(b) = 2 \int_b^{\Rs} \frac{n_3(r)\sigma_{\lambda}\Phi(\lambda)r}{\sqrt{r^2-b^2}} dr
    \label{eq:tau}
\end{equation}
where $n_3$ is the number density of the metastable triplet helium, $\sigma_{\lambda}$ is the absorption cross section, and $\Phi(\lambda)$ is the Voigt line profile($\int_0^{\infty} \Phi(\lambda)d\lambda = 1$).
The absorption depth mainly depends on the fraction of the helium triplet. In metal-rich atmospheres, metal cooling reduces the mass-loss rate and the gas temperature.

Approximating the upper region of the planetary atmosphere as an isothermal hydrostatic layer, the number density can be given by
\begin{equation}
    n_3 = n_{\rm base} f_3\exp\left(-\frac{G\Mp}{\Rp\cs}+\frac{G\Mp}{\cs^2r}\right)
    \label{eq:density_profile}
\end{equation}
where $f_3$ is the fraction of metastable helium which depends on the atmospheric temperature. We note that the fraction weakly depends on the altitude \citep{Oklopcic_2019} and can be assumed constant. 
Eq.~\ref{eq:tau} can be rewritten as:
\begin{equation}
    \tau_{\lambda}(b) = A\int_b^{\Rs} \frac{r\exp(a/r)}{\sqrt{r^2-b^2}}dr
    \label{eq:tau2}
\end{equation}
 The absorption cross section is 
\begin{equation}
    \sigma_{\lambda} =\frac{\pi e^2\lambda_0}{m_ec^2}f
\end{equation}
where f is the oscillator strength of the transitions from NIST database. We use three helium metastable lines (10830.34\,\AA, 10830.25\,\AA, 10829.09\,\AA ).

We introduced two parameters for the analytical expression of the equivalent width.
 \begin{equation}
        a =\frac{G\Mp}{\cs^2},\  A = 2n_{\rm base} f_3 \sigma_{\lambda}\exp\left(-\frac{G\Mp}{\Rp\cs^2}\right)\Phi(\lambda)
 \end{equation}
where the base density $n_{\rm base} = \sqrt{\Phi_{\rm EUV}/\alpha_B/H_0}$, $\Phi_{\rm EUV}$ is the EUV photon flux at the planet, $\alpha_B$ is the recombination coefficient of hydrogen atom, and $H_0=\cs^2\Rp^2/2G\Mp = \Rp^2/a$ is the pressure scale height for an isothermal sound speed $\cs$.

\subsection{The case of high-mass planets}
\label{app:derivation_high}
To simplify Eq.~\ref{eq:tau2}, we use the substitute of variables $u=\sqrt{r^2-b^2}, dr = udu/\sqrt{u^2+b^2}$. And if $a$ is larger than $b$ (high-mass planets), the contribution from $u\sim0$ region dominates the integral. The optical depth in Eq.\ref{eq:tau2} is given as:
\begin{equation}
    \begin{split}
        \tau_{\lambda}(b) &= A\int_0^{\sqrt{\Rs^2-b^2}}\exp\left(\frac{a}{\sqrt{u^2+b^2}}\right)du\\
        &\sim A\int_0^{\infty} \exp\left(\frac{a}{\sqrt{u^2+b^2}}\right)du\\
        &\sim A\exp\left(\frac{a}{b}\right)\int_0^{\infty}\exp\left(-\frac{au^2}{2b^3}\right)du\\
        & = A\sqrt{\frac{\pi b^3}{2a}}\exp\left(\frac{a}{b}\right)
    \end{split}
    \label{eq:tau_high_mass}
\end{equation}

In the optically thin limit, $\exp(-\tau)\sim 1-\tau$ and the transmission spectrum in Eq.~\ref{eq:transmission} can be approximated as:
\begin{equation}
        T_{\lambda} \sim 1 -  \frac{2A\sqrt{\frac{\pi}{2a}}}{(\Rs^2-\Rp^2)}\int_{\Rp}^{\Rs}  b^{5/2}\exp\left(\frac{a}{b}\right) db
\end{equation}

We compute the second term using Wolfram Alpha and get
\begin{equation}
\begin{split}
      I_1(b) &= \int b^{5/2}\exp\left(\frac{a}{b}\right) db\\
      &= \frac{2}{105}b^{3/2} (4a^2 + 6ab + 15b^2)e^{a/b}\\
      &+\frac{8}{105}i a^{7/2}\Gamma(-1/2,-a/b) + \mathrm{constant}
\end{split}
\end{equation}
where $\Gamma(s,x)$ is the incomplete gamma function. 
Finally, by integrating the transmission, the equivalent width of He I absorption can be given as:
\begin{equation}
    W_\lambda = \frac{2A'\sqrt{\frac{\pi}{2a}}}{(\Rs^2-\Rp^2)} (I_1(\Rs)-I_1(\Rp))
    \label{eq:EW_1}
\end{equation}
where $A' = A/\Phi(\lambda)$. 
Note that $I(b)$ includes an imaginary part but the equivalent width only has a real value because the imaginary part will be canceled out in Eq.~\ref{eq:EW_1}.

\subsection{The case of low-mass planets}
\label{app:derivation_low}
As in \cite{Zhang_2023c}, the equivalent width of helium is roughly proportional to the mass-loss rate in sub-Neptunes.  
In the case of close-in low-mass planets (e.g., sub-Neptunes), $a$ is no longer large enough that we can assume the upper atmosphere to be hydrostatic because the gas becomes supersonic around the planet (within a few $\Rp$) and the gas velocity is almost constant in close-in exoplanets. In this case, the density profile can be approximated $\rho(r)\sim\rho_{\rm base}(\Rp/r)^{\beta}$. We first tested the simple constant velocity case with $\beta=2$
and the optical depth in Eq.~\ref{eq:tau} becomes:
\begin{equation}
    \begin{split}
    \tau_{\lambda}(b)&\sim B\int_b^{\Rs} \frac{1}{r\sqrt{r^2-b^2}} dr\\
    &=B\frac{\arctan(\frac{\sqrt{\Rs^2-b^2}}{b})}{b}
    \end{split}
\end{equation}
where 
\begin{equation}
    B=2n_{\rm base}f_3\Rp^2\sigma_{\lambda}\Phi(\lambda)
\end{equation}
In the optically thin limit, the transmission spectrum can be given as:
\begin{equation}
    T_{\lambda} \sim 1 - \frac{2B}{\Rs^2-\Rp^2}\left(\Rp\arctan\left(\frac{\sqrt{\Rs^2-\Rp^2}}{\Rp}\right) - \sqrt{\Rs^2-\Rp^2}\right)
\end{equation}
from the fact that:
\begin{equation}
    \int db \arctan{\frac{\sqrt{r^2-b^2}}{b}} = b\arctan{\frac{\sqrt{r^2-b^2}}{b}}-\sqrt{r^2-b^2} +\mathrm{constant}
\end{equation}

For low-gravity planets, the effective planetary radius is larger than the planetary radius. We calculate the effective radius from 
\begin{equation}
R_{\rm EUV} = \Rp \left(1+\frac{\Rp}{2 R_{\rm B}} \log(n_{\rm base}/n_{\rm surf})\right)^{-1}
\label{eq:R_euv}
\end{equation}    
where $R_{\rm B}=G\Mp/2c_{\rm surf}^2$ is the Bondi radius with sound speed on the planetary surface and $n_{\rm surf}$ is the surface number density. We define the surface as the inner boundary of the simulations ($r=\Rp$). 
If the effective radius is larger than the sonic point radius, we treat the effective radius as the sonic point radius. Note that such a condition is only satisfied in very low gravity planets as in \cite{Mitani_2025}. We also test $\beta=3$ which is more consistent with the density profile from our hydrodynamics simulations because of the acceleration outside the sound speed point. 
In the case of $\beta=3$, 
\begin{equation}
    \begin{split}
    \tau_{\lambda}(b)&\sim B'\int_b^{\Rs} \frac{1}{r^2\sqrt{r^2-b^2}} dr\\
    &=B'\frac{\sqrt{\Rs^2-b^2}}{b^2\Rs}
    \end{split}
\end{equation}
where 
\begin{equation}
    B'=2n_{\rm base}f_3\Rp^3\sigma_{\lambda}\Phi(\lambda)
\end{equation}
and the transmission spectrum can be given as:
\begin{equation}
    T_\lambda \sim 1 + \frac{2B'}{\Rs^2-\Rp^2}\left( \ln\tan\theta/2 + \cos\theta \right)
\end{equation}
where 
\begin{equation}
    \theta = \arcsin\left(\frac{\Rp}{\Rs}\right).
\end{equation}

\subsection{Fitting of analytic model}
For high-mass planets, our analytic formula contains the incomplete gamma function. For practical use, we fit the incomplete gamma function by simpler forms.
We test three types of fitting:
\begin{equation}
    \begin{split}
        f_1(x) &= \alpha_1x^{\alpha_2}\\
        f_2(x) &= \beta_1x^{\beta_2}e^{\beta_3 x} \\
        f_3(x) &= \gamma_1 e^{\gamma_2 x} +  \gamma_3 e^{-\gamma_4 x} +  \gamma_5 e^{x} / x^{1.5} 
    \end{split}
\end{equation}

\begin{figure}
    \centering
    \includegraphics[width=1.\linewidth]{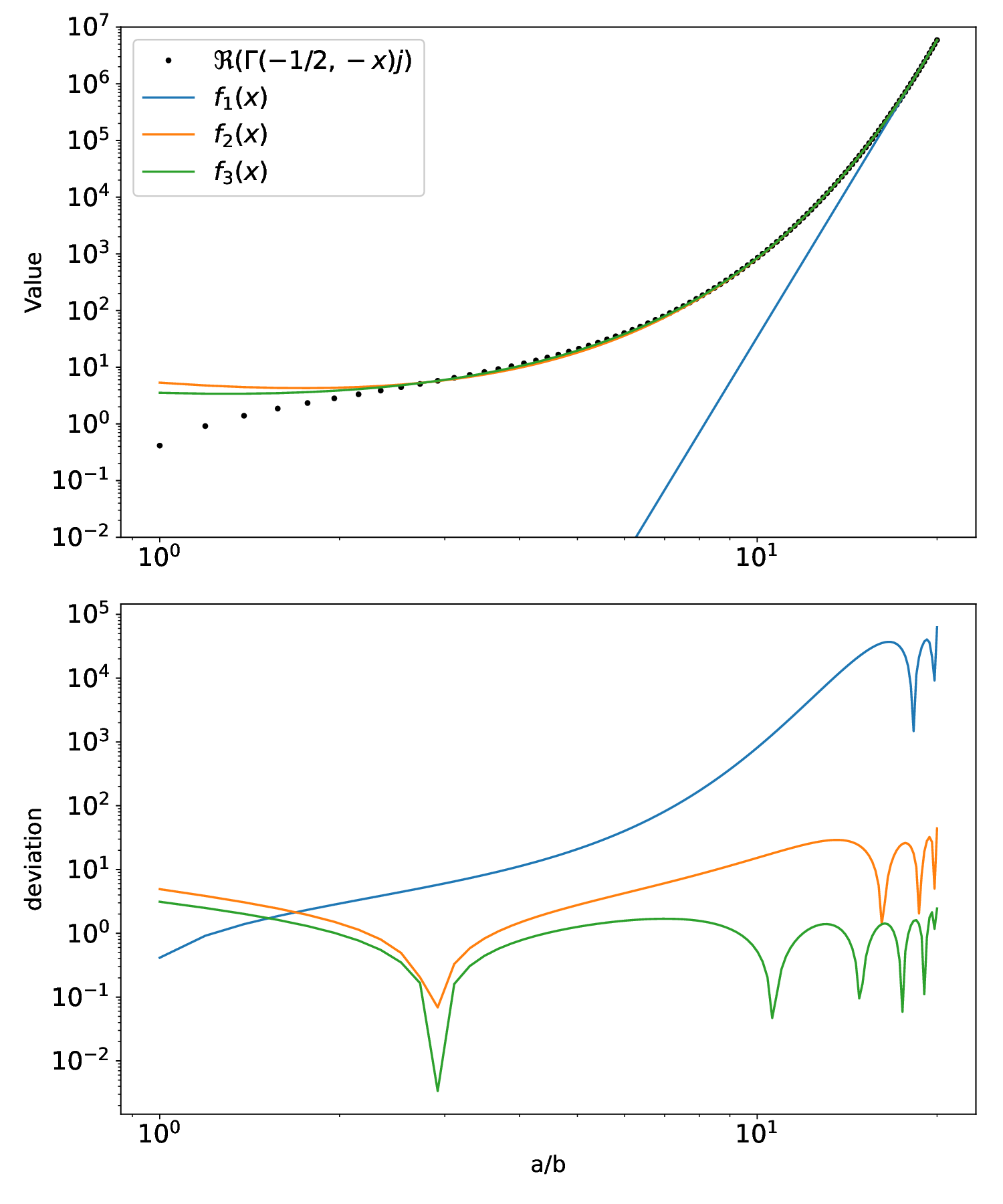}
    \caption{Upper panel: the real part of the incomplete gamma function $\Gamma(-1/2,-a/b)i$ (black points) with fitting with $f_1(x)$(blue line), $f_2(x)$(orange line), and $f_3(x)$ (green line). Lower panel: deviations from the incomplete gamma function.  }
    \label{fig:fit_gamma}
\end{figure}
We show the fitting of the function in Figure~\ref{fig:fit_gamma}. We find that the simple power law $f_1(x)$ is not enough to fit the function. The fitting with power law and exponential function seems to be well-fitted but we find that $f_2(x)$ cannot reproduce the equivalent width because of the relatively small deviation at high x value. In the fitting with $f_3(x)$, we add the asymptotic function and confirm that the $f_3(x)$ can reproduce the equivalent width with the exact incomplete gamma function.  
The fitting parameters for $f_3(x)$ are $\gamma_1 = 0.019607711263469527, \gamma_2 = 0.8094995623133185, \gamma_3 = 0.39743620918312733, \gamma_4 = -0.4886404403069673,\gamma_5 = 1.0464152457944846$.

\section{One-dimensional radiation-hydrodynamics simulations with metal cooling}
\label{sec:method}
The temperature of the planetary upper atmosphere depends on many stellar/planetary parameters. To test our analytical model, we performed one-dimensional radiation-hydrodynamics simulations.
In this section, we introduce the metal cooling dependence of the atmospheric temperature and the equivalent width of helium absorption.
We used the ATES-code \citep{Caldiroli_2021} to calculate the one-dimensional (1D) atmospheric profile including a self-consistent helium triplet population. We set the extreme ultraviolet (EUV) spectra to 100 times that of GJ 876 spectra from the MUSCLE treasury survey \citep{Loyd_2016} and the far ultraviolet (FUV) to the same value. 
This choice is intended to mimic systems in which the He I 10830 \AA\, line is most easily detectable, i.e. with strong EUV-driven escape and relatively weak FUV photoionization from the metastable level. At fixed EUV flux, increasing the FUV flux would enhance photoionization of metastable helium and thus reduce the metastable fraction and the resulting equivalent width at a given mass-loss rate, whereas lowering the EUV flux would generally decrease the mass-loss rate and metastable He density. A comprehensive exploration of the full EUV-FUV parameter space is left for future work.
For low-mass host stars with weak FUV flux, the metastable helium fraction is almost independent of FUV flux \citep{Oklopic_2019}. Our model is valid for young close-in planets around low-mass stars with strong EUV and weak FUV flux. We confirmed that the equivalent width of helium absorption is almost the same even if we assume 100 times FUV flux.

\subsection{Basic equations}
We solved the following hydrodynamic equations using the ATES code:
\begin{align}
&\frac{\partial \rho}{\partial t}+\frac{1}{r^2}\frac{\partial}{\partial r} (r^2\rho v) =0\\
&\frac{\partial\rho v}{\partial t} + \rho v\frac{\partial v}{\partial r} = -\frac{\partial p}{\partial r} -\rho \frac{\partial \Psi}{\partial r}\\
&\frac{\partial E}{\partial t} + \frac{\partial Hv}{\partial r} = -\rho v \frac{\partial \Psi}{\partial r} + \rho (\Gamma-\Lambda)
\end{align}
where $\rho, v, p, E$, and $H$ are the gas density, velocity, pressure, energy, and enthalpy per unit volume of gas. The effective gravitational potential $\Psi$ can be given by $\Psi=-G\Mp/r-GM_*/r_*-GM_*r_*^2/2l^3$ where $l,r_*$ represent the semimajor axis and local distance to the host star as in \cite{Mitani_2022}. We used the EUV photoionization heating rate in $\Gamma$ and the Ly$\alpha$ cooling and metal cooling in $\Lambda$. The code and setup were already tested in \citet{Mitani_2025}. We set the surface density as $n(r=\Rp)=10^{14}\mathrm{\, cm^{-3}}$ and the temperature $1000 \rm{\,K}$ at the inner boundary.

\subsection{Metal cooling effect}
We incorporate Mg cooling rate into the total cooling rate ($\Lambda$) in our simulations. Mg cooling is a dominant cooling source in the upper atmosphere of hot Jupiters \citep{Huang_2023,Fossati_2025}. 

\begin{equation}
    \Lambda_{\rm Mg}(\Phi_{\rm EUV}, \Mp, \Rp) = n_{\rm base} f_{\rm Mg} \frac{\Delta E}{n_l/n_u} A_{ul} \frac{n_e}{n_{\rm crit}}
\end{equation}
where $f_{\rm Mg}$ is the fraction of MgII, $\Delta E = 4.4\mathrm{\,eV}$, $n_l,n_u,n_{\rm crit},A_{ul}$ are the lower-, upper-level, critical density, and the Einstein coefficient. The Mg cooling can be a dominant cooling process in the upper atmosphere \citep{Huang_2023}. We note that the cooling rate of Mg I and Mg II is similar and our model with Mg II cooling can be applied for planets with Mg II and without Mg I. 

We also investigated oxygen cooling to better understand the atmospheres of relatively cool planets where Mg condenses out of the upper atmosphere. Fine structure transitions can serve as a dominant cooling mechanism in such environments. For oxygen cooling, we considered the energy levels $^3P_2$, $^3P_1$, $^3P_0$, $^1D_2$, and $^1S_0$. Assuming statistical equilibrium, we calculated the level populations using collisional excitation and de-excitation rates caused by hydrogen and electrons, as well as Einstein coefficients \citep{1989_Osterbrockbook,1989_HollenbachMcKee}.

However, other metal species play an important role in the atmosphere of sub-Neptunes \citep{Zhang_2022,Linssen_2024,Kubyshkina_2024}.  
Interestingly, the cooling rate of iron is comparable to that of magnesium, playing a significant role in metal-rich planetary atmospheres. Our model indicates that efficient metal cooling, such as from Mg and Fe, can reduce atmospheric mass loss by lowering temperatures and influencing escape dynamics.

Our analytic formula of the equivalent width does not contain any assumptions on the metal species and the gas temperature determines the results. The formula can be used to understand the temperature of the observed upper atmosphere.
We note that the helium-to-hydrogen number fraction in our simulations is fixed at $0.083$. This assumption is reasonable for planets with not extremely metal-rich upper atmospheres ($Z < 50~Z_{\odot}$). However, the helium fraction remains one of the most significant sources of uncertainty in our model. In the upper atmosphere, the metallicity is expected to be relatively lower compared to the lower atmosphere due to the condensation of metal species and our assumption is valid for the majority of planets with hydrogen-dominated atmospheres.

In low-temperature planets, Mg atoms can condense into clouds in the lower atmosphere, reducing the abundance of Mg in the upper atmosphere. The condensation temperature of Mg species is approximately $2000 ~\mathrm{ K}$, so our model, which includes Mg cooling, is applicable to close-in planets with high surface temperatures. Notably, solar-abundance magnesium has been detected in the upper atmosphere of planets with surface temperatures around 1500~K (e.g., HD 209458 b, \cite{Vidal-Madjar_2013}).

\begin{figure}[h]
    \centering
    \includegraphics[width=1\linewidth]{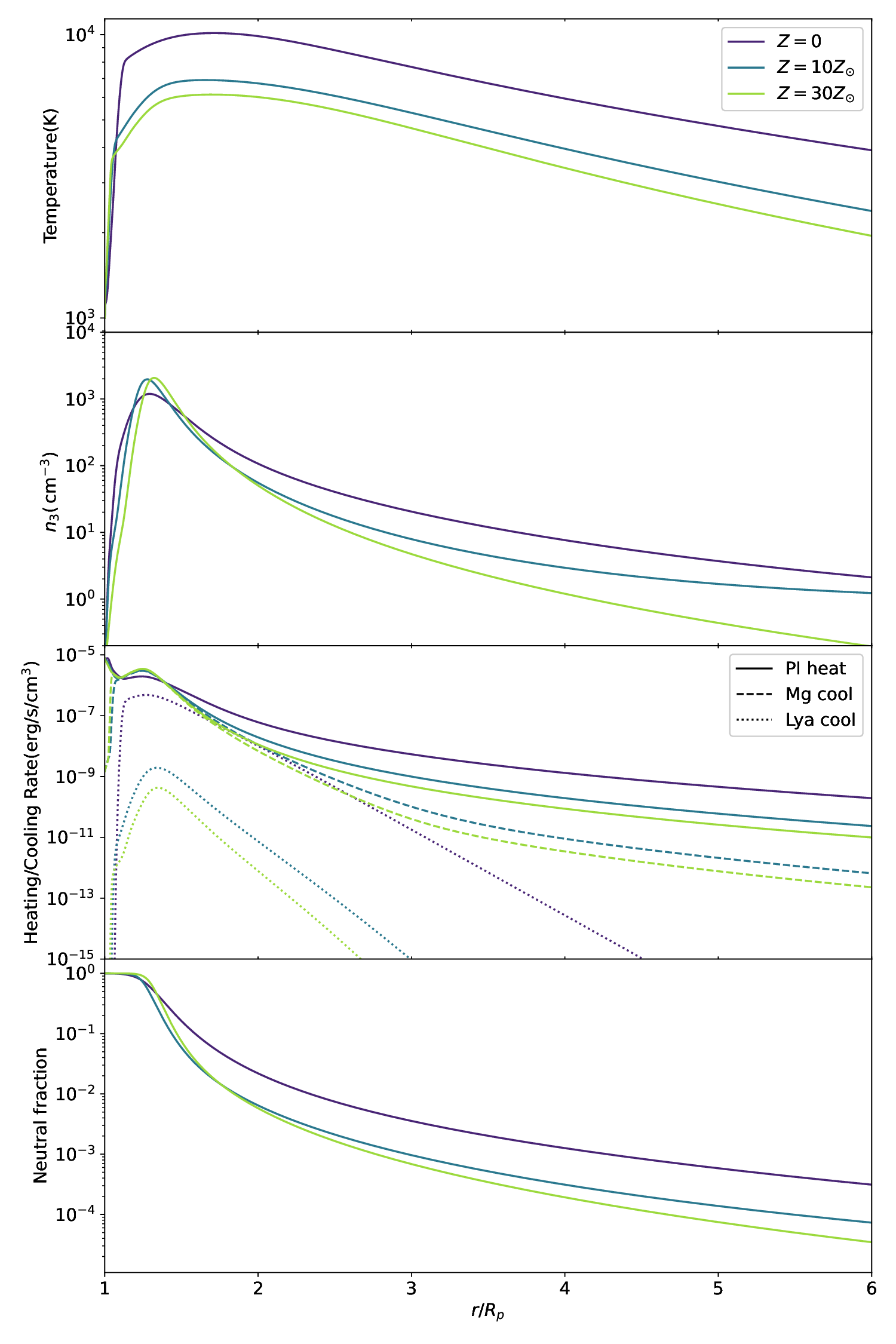}
    \caption{The radial profiles of temperature (top), metastable helium density(upper-middle), and heating/cooling rate(lower-middle), neutral fraction ($n_{\rm neutral}/n_{\rm total}$ bottom) with high-mass planets different metallicity ($\Mp=0.7\MJ,\Rp=1.4\RJ, l=0.05\,\mathrm{au}, F_{\rm EUV} = 5600\mathrm{\, erg/s/cm^2}, Z=0,10~Z_{\odot},30~Z_{\odot}$). The photoionization heating (solid), Mg cooling (dashed), and Ly$\alpha$ cooling (dotted) are shown in the bottom panel.
    }
    \label{fig:profile_HJ}
\end{figure}

\begin{figure}[h]
    \centering
    \includegraphics[width=1\linewidth]{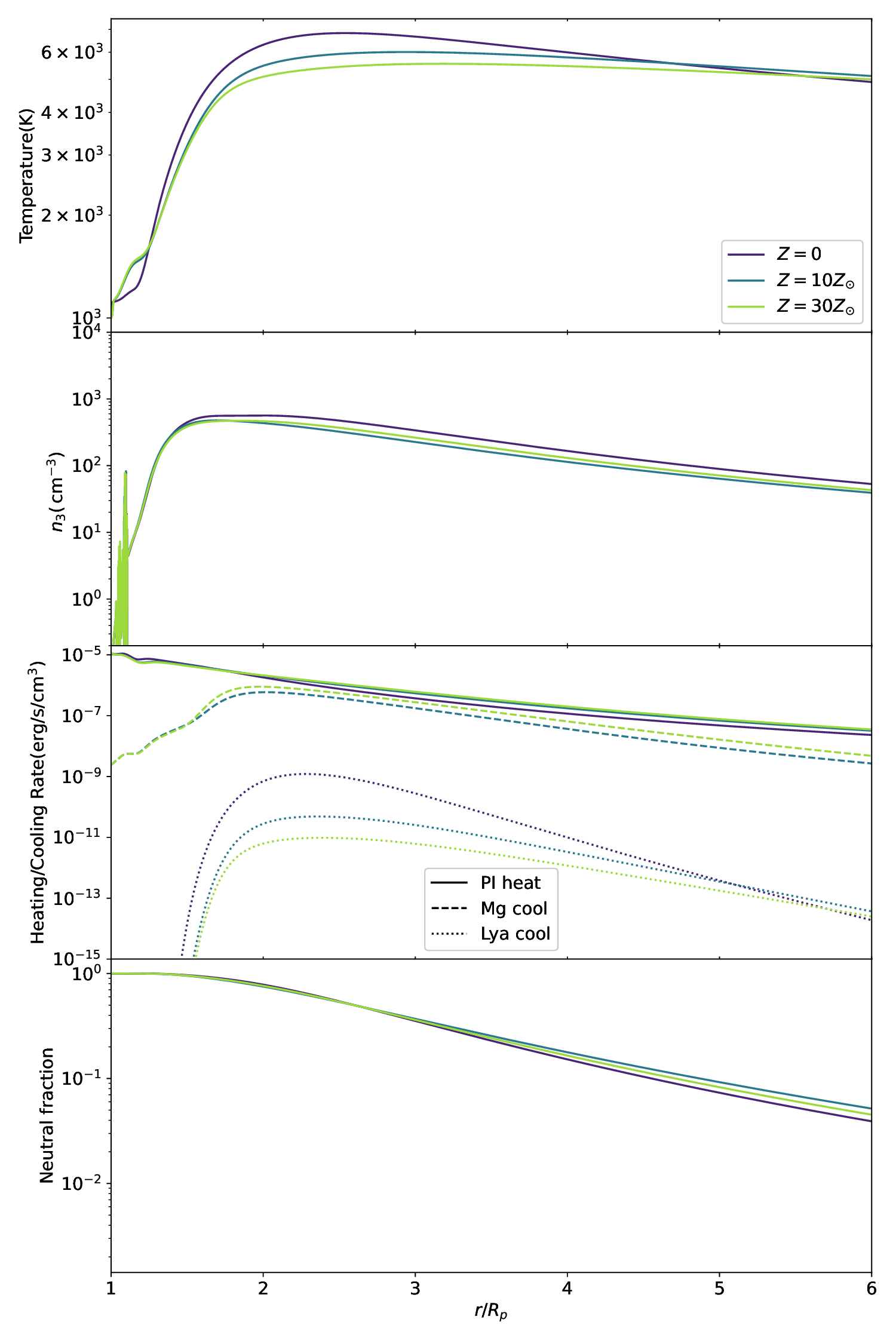}
    \caption{Same as Fig.~\ref{fig:profile_HJ}, but for low-mass planets ($\Mp=10.2~\Mearth,\Rp=0.25~\RJ, l=0.05\,\mathrm{au}, F_{\rm EUV} = 5600\mathrm{\, erg/s/cm^2}$).
    }
    \label{fig:profile_SN}
\end{figure}

We assume that the metal composition is similar to the solar value and ran simulations of a gas giant ($\Mp=0.7\MJ,\Rp=1.4\RJ$, $l=0.05\,\mathrm{au}$) as the fiducial cases to test the planet like HD209458b which has been already observed in helium. We present the radial profiles of high-mass planets in Figure~\ref{fig:profile_HJ}. Due to Mg cooling, the temperature of metal-rich atmospheres decreases. The metastable helium density remains similar across different metallicities because lower temperatures lead to a higher fraction of helium in the metastable state. 
We also present the hydrodynamics profiles of low-mass planets ($\Mp=10.2~\Mearth,\Rp=0.25~\RJ, l=0.05\,\mathrm{au}$) like HD 73583b,  \citep{Zhang_2022} in Figure~\ref{fig:profile_SN}.
We also ran simulations with only oxygen line cooling and found that oxygen cooling is negligible for low-metallicity cases ($Z < 50~Z_{\odot}$). This result is consistent with the previous simulation with oxygen cooling \citep{Zhang_2022}. 
To test our analytical model, we focused on Mg cooling to understand helium absorption in metal-rich planets. 

Previous studies \citep{Zhang_2022,Zhang_2023c,Orell-Miquel_2024} used a rather simple relationship between the mass-loss rate and the equivalent width to estimate the mass-loss rate:
\begin{equation}
    \dot{M} = \frac{R_*m_em_{\rm He}\cs c^2 }{0.25 f_3 e^2 \lambda_0^2\Sigma g_l f} W 
    \label{eq:Mdot_EW}
\end{equation}
where $\cs$ is the sound speed, $e$ and $m_e$ are the electron charge and mass, $c$ is the speed of light, $\lambda_0=10830$\,\AA, and $W$ is the equivalent width. $g_l$ and $f$ represent the degeneracy and oscillator strength. We adopted $\Sigma  g_l f=1.62$ as previously used in \cite{Zhang_2022}. In the derivation of Eq.~\ref{eq:Mdot_EW}, they assumed optically thin limit and the timescale of the replacement of helium triplet $\tau=R_*/\cs$. In their estimate, the radial profile of density is neglected.
Constructing a model for equivalent width with the 1D radial profile is the main goal of this section.

Note that the previous study suggests the metal line absorption of near-Ultraviolet (NUV) is a dominant heating source in the upper atmosphere of ultra-hot Jupiter \citep{Fossati_2025}. In this study, we ignore the heating due to the metal absorption because such a cooling is significant only for planets around A-type stars with extreme intense NUV luminosity due to the high surface temperature and negligible for many planets.

\subsection{Post-processing other metal cooling effects}
In real atmospheres, radiative cooling by multiple metal species can further reduce the temperature, as demonstrated in previous studies \citep{Zhang_2023c, Linssen_2024, Kubyshkina_2024, Fossati_2025}. To assess this effect, we compute the metal cooling rate in post-processing using the \texttt{sunbather} module \citep{Linssen_2024}, which is based on the open-source code CLOUDY \citep{Chatzikos_2023}. We use the hydrodynamics profile from our simulations and calculate the heating and cooling rate by \texttt{sunbather} module to check how the detailed metal coolings change the atmospheric temperature structure. 
We note that our analytical models assume the isothermal profile of atmosphere and isotropic helium metastable state fraction and are independent of the dominant cooling process.

\subsection{Thermospheric balance temperature with metal cooling}
If the EUV flux is sufficiently large, the gas temperature reaches high enough level that the radiative cooling dominates the cooling process. We estimate the gas temperature where the radiative cooling balances the photoionization heating at the base as in \cite{Mitani_2025}:
\begin{equation}
    \Gamma_{\rm EUV} = \Lambda_{\rm metal}
    \label{eq:T_th}
\end{equation}

\begin{figure}[h]
    \centering
    \includegraphics[width=1\linewidth]{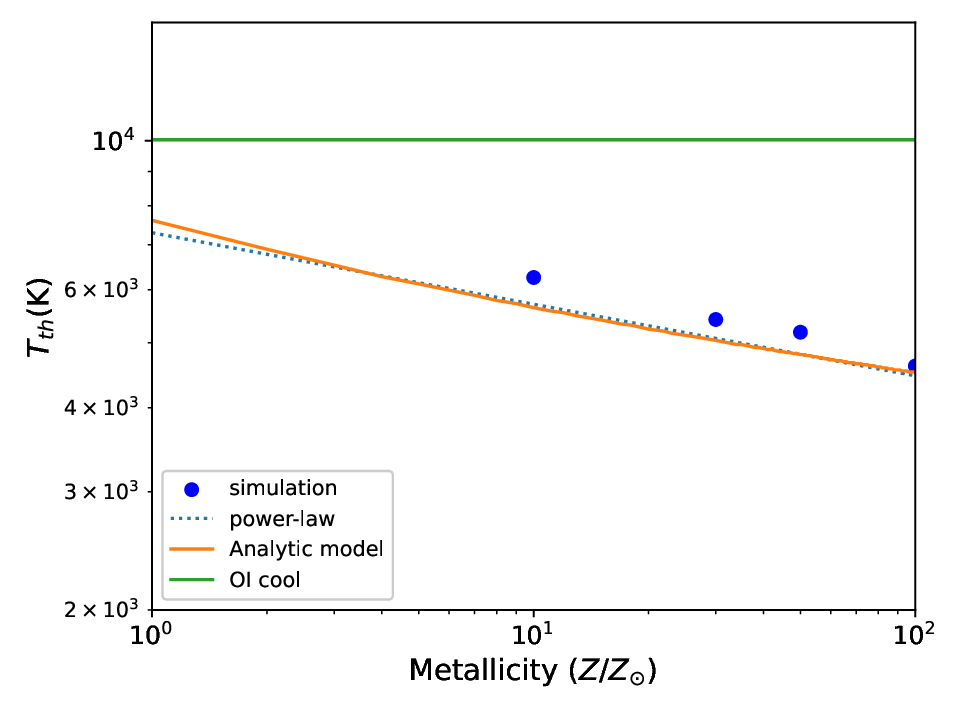}
    \caption{Thermospheric balance gas temperatures as a function of metallicity. The temperature from Eq.~\eqref{eq:T_th} (solid curve, orange; Mg cooling, green; OI cooling) and gas temperature at the sonic point from simulations (dots) are shown. The power-law fit of \eqref{eq:Teqfit} is also shown as a dotted curve.  }
    \label{fig:T_th_Z}
\end{figure}

We assume the monochromatic EUV ($h\nu_1=20\mathrm{\, eV}$) to estimate the base density where the outflow launched by EUV photoheating for simplicity. Note that the base density depends on the EUV spectrum within a factor of two as in \cite{Mitani_2025}.
Figure~\ref{fig:T_th_Z} shows the metallicity dependence of the gas temperature.
Our simple estimate from Eq.~\eqref{eq:T_th} well represents the metallicity dependence of gas temperature qualitatively. 
We find that the metallicity dependence of the temperature can be given by a power-law. We fit the metallicity dependence of the thermospheric balance temperature in the range $Z_{\odot} \le Z \le 100~Z_{\odot}$ and get:
\begin{equation}
\label{eq:Teqfit}
    T_{\rm th} = T_0 \left(\frac{Z}{Z_{\odot}}\right)^{\beta}
\end{equation}
where $T_0 = 7562\mathrm{\, K} , \beta = -0.1146$.  We note that the gas thermospheric balance temperature is determined by Ly$\alpha$ cooling and becomes $\sim 10^4{\, \rm K}$ for planets without metal cooling.
We also confirm that the metallicity dependence of the thermospheric balance temperature is similar in weak-gravity planets.

\section{Comparison of analytical model and simulation}
In this section, we compare the analytical model described in Section~\ref{sec:He_analytic} with our simulations described in Section~\ref{sec:method}. We also compare our analytical model with the previous relationship between mass loss and equivalent width. Eq.~\ref{fig:Mdot_EW}. 
\subsection{High-mass planets}
\begin{figure}
    \centering
    \includegraphics[width=1\linewidth]{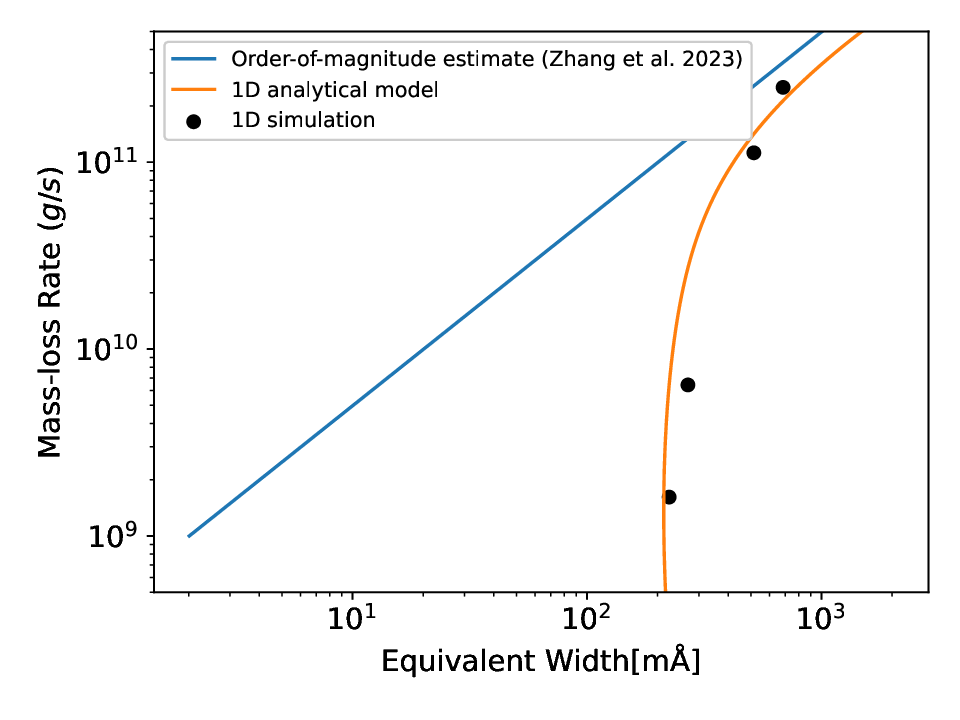}
    \caption{Relationship between equivalent width and mass-loss rate. The solid blue curve represents the order-of-magnitude estimate by \citet{Zhang_2023c}. The solid orange curve represents our 1D analytical model. The results of our 1D simulations with different metallicity ($Z=0,1,10,30\, Z_{\odot}$) are also shown as black dots.}
    \label{fig:Mdot_EW}
\end{figure}

In case of atmosphere irradiated by hard EUV such as M-stars, the fraction of He triplet is approximated as in \cite{Oklopcic_2019}:
\begin{equation}
    f_3\sim7\times 10^{-6}\left(\frac{10^4\mathrm{\,K}}{{\rm T_{\rm th}}}\right)^{0.8}(1-e^{-\Phi_1 \Rp/c_s})
    \label{eq:He_fraction1}
\end{equation}
where $\Phi_1$ is the photoionization rate of the helium ground state.
In the case of high-mass planets, the sound crossing timescale $\tau_{\rm adv} = \Rp/\cs $ is longer than the r
photoionization timescale of the helium ground state and the last term in Eq.~\ref{eq:He_fraction1} can be neglected.
We performed 1D simulations of hot Jupiters ($\Mp=0.7\MJ, \Rp=10^{10}\mathrm{\,cm}$) with different metallicity. 
We also calculate the equivalent width from Eq.~\ref{eq:EW_1} with different thermospheric balance temperatures and sound speed $\cs = \sqrt{2k_B T_{\rm th}/m_{\rm H}}$.
Figure~\ref{fig:Mdot_EW} shows the derived equivalent width with the mass-loss rate $\dot{M} = \pi\Rs^2 \rho_{\rm s} \cs$. We also compare our analytical model to the previous order-of-magnitude estimate in Eq.~\ref{eq:Mdot_EW} and the simulation results.

Our simulations and analytical model suggest the existence of a lower limit for the equivalent width. The low temperature leads to a low-density profile and a high fraction of He triplet, as described in Eq.~\ref{eq:He_fraction1}. These two effects almost cancel each other out, making the equivalent width almost independent of metallicity (temperature).
For hot Jupiters exposed to high EUV flux, order-of-magnitude estimates may overestimate the mass-loss rate. Estimating the mass-loss rate of hot Jupiters based on helium triplet equivalent width is challenging because the equivalent width depends only weakly on the mass-loss rate.

\subsection{Low-mass planets}
\begin{figure}
    \centering
    \includegraphics[width=1\linewidth]{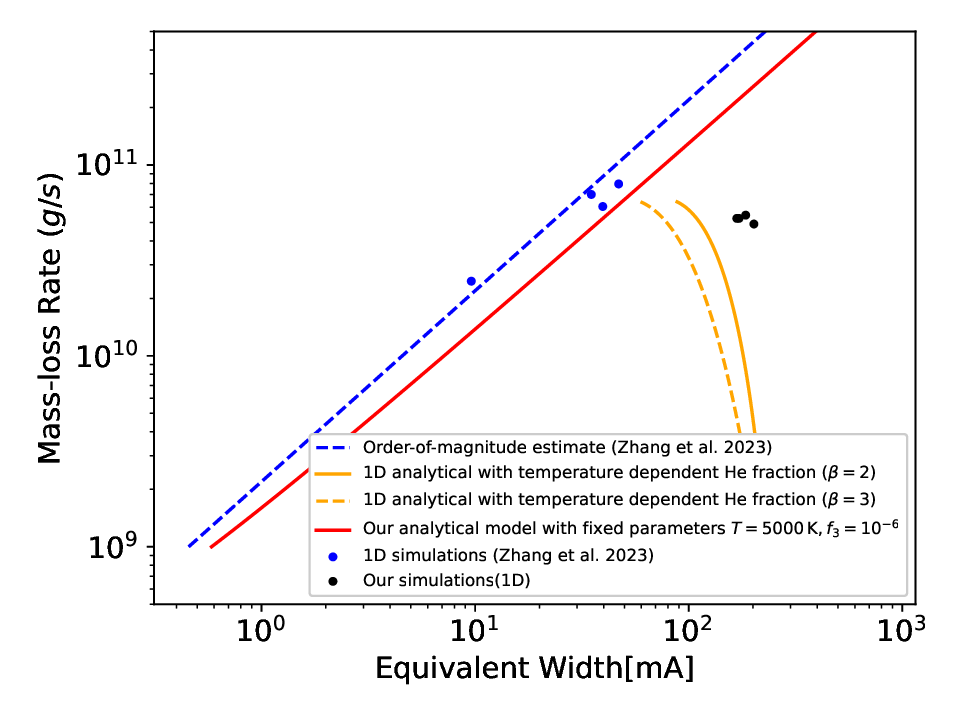}
    \caption{Same as Figure~\ref{fig:Mdot_EW} but for low-mass planets. Solid curves represent our 1D analytical model (orange: with temperature dependent helium metastable fraction; red: with fixed temperature and helium metastable fraction). The results of 1D simulations ($Z=0,1,10,30\, Z_{\odot}$) are shown as dots (black: our simulations; blue: from \cite{Zhang_2023c}). }
    \label{fig:Mdot_EW_lowa}
\end{figure}

For low-mass planets with small $a$, the equivalent width can be given as (see \ref{app:derivation_low}):
\begin{equation}
    W_{\lambda} \sim 1 - \frac{2B}{\Rs^2-\Rp^2}\left(\Rp\arctan\left(\frac{\sqrt{\Rs^2-\Rp^2}}{\Rp}\right) - \sqrt{\Rs^2-\Rp^2}\right)
    \label{eq:EW_2}
\end{equation}
In low-mass planets with low UV flux, the effect of advection cannot be neglected and the last term in Eq.~\ref{eq:He_fraction1} becomes dominant and we take into account the effect of advection as in \citet{Oklopcic_2019}.
We also calculate the equivalent width for low-mass planets (HD 73583b, $\Mp=10.2~\Mearth, \Rp=0.25~\RJ, R_* = 0.63~\Rsun, F_{\rm EUV} = 5600\mathrm{\, erg/s/cm^2}$) with different temperatures ($3000\mathrm{\, K}<T<15000\mathrm{\, K}$). 
The ionization fraction of hydrogen is not large at low altitudes and we can assume there the mean molecular weight $\mu \approx 1$. 
Figure~\ref{fig:Mdot_EW_lowa} shows our analytical model and the results of different simulations. Our analytical model suggests an equivalent width that is insensitive to the mass-loss rates.
We find that the equivalent width from our simulations is larger than that from the previous study because the helium metastable fraction of the previous study is lower than ours. This is because the high temperature in the upper atmosphere of the previous study leads to a high rate of recombination into the triplet state. 
The density profile difference in our analytical models is within a factor of 2 in the case of temperature dependent metastable helium fraction. Our analytical method is valid for general planets.
To understand the relationships between the mass-loss rate and the equivalent width, we tested our models with fixed parameters ($T=5000\mathrm{\,K}, f_3=10^{-6}$) and found that our analytical model reproduces the behavior of the previous study under this assumption (red solid line in Fig.~\ref{fig:Mdot_EW_lowa}).

The previous estimate (Eq.~\ref{eq:Mdot_EW}) is better suited for low-mass planets than for high-mass planets because the density structure of low-mass planets is not steep and the radial dependence of the optical depth is not significant.
\section{Discussions}
\label{sec:discussion}

\subsection{Model predictions and observations}
Helium triplet absorption has emerged as a crucial probe of the upper atmospheres of close-in exoplanets. Although several recent observations have reported non-detections of this feature, our results can be used to understand the atmospheric properties of such observed planets. Our radiation-hydrodynamics model, which explicitly includes metal cooling, demonstrates that enhanced metal radiative cooling can lower the atmospheric temperature. This reduction, in turn, increases the fraction of helium in its metastable triplet state, offsetting the expected decrease in mass-loss rate.
We have analyzed the equivalent width of helium triplet absorption using data from the MOPYS project \citep{Orell-Miquel_2024}.
\begin{figure}
    \centering
    \includegraphics[width=1.\linewidth]{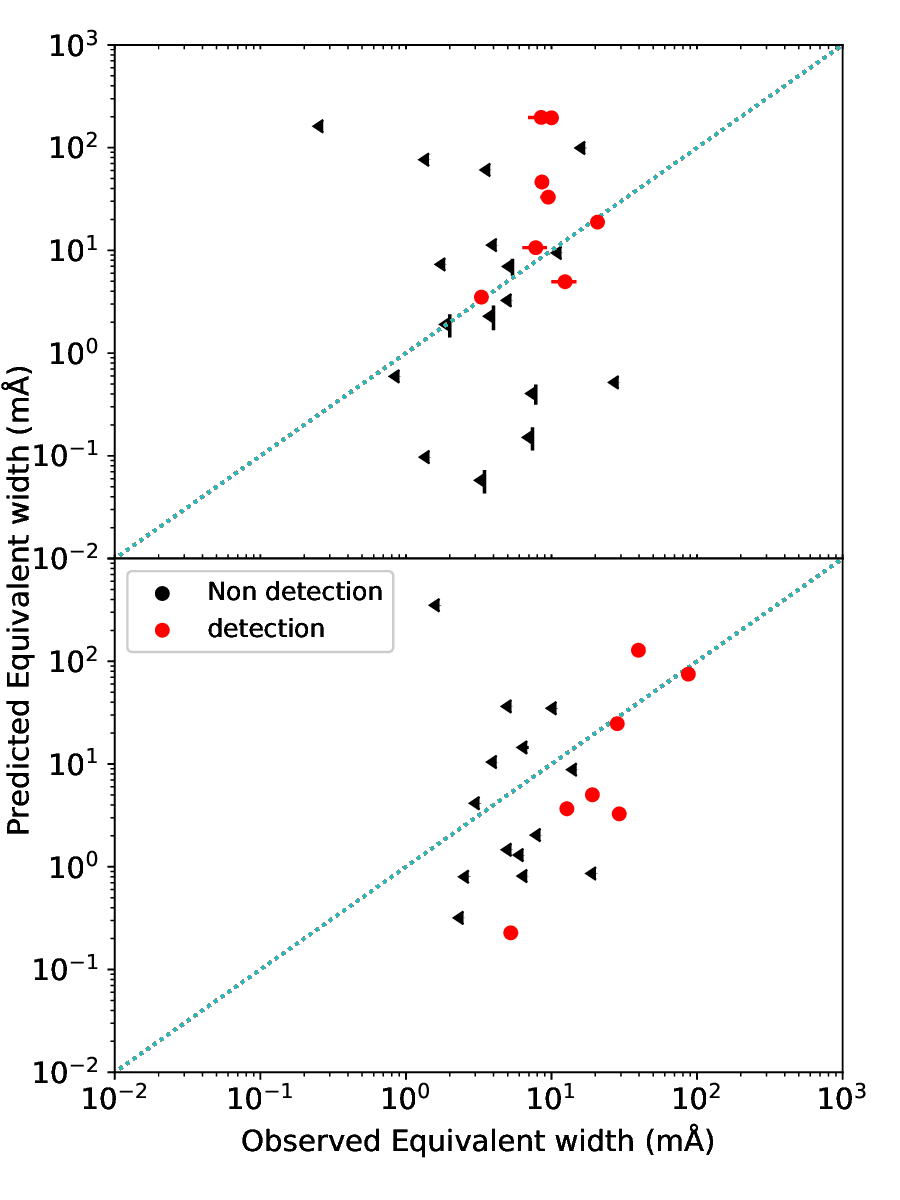}
    \caption{Observed equivalent width and prediction of the equivalent width from our 1D model. Low-mass planets ($\Mp\le0.1M_{J}$, top), High-mass planets ($\Mp>0.1M_{J}$, bottom). We use the observational data from MOPYS project \citep{Orell-Miquel_2024}. For planets with non-detection, the triangles show the upper limit of the equivalent width. }
    \label{fig:observation}
\end{figure}

In Figure~\ref{fig:observation}, we present the distribution of estimated equivalent widths of exoplanets with helium triplet observations for low-mass planets. We assume a gas temperature of $T_{\rm gas}=5000\mathrm{\, K}$ because the metallicity of the atmosphere is usually unknown and the equivalent width is almost independent of the gas temperature.  
Planets with detection of helium triplet show larger equivalent width ($>10$ m\AA) in Eqs.~\ref{eq:EW_1} and \ref{eq:EW_2} than that of planets without detection of helium absorption.
Our findings suggest that in cases where the observed upper limit of the equivalent width exceeds our prediction, the actual helium triplet fraction may be lower than estimated by Eq.~\ref{eq:He_fraction1}. Such a reduction in the triplet fraction could result from additional processes, such as FUV photoionization of metastable helium or the influence of stellar winds, as discussed in \cite{Mitani_2022}. In Figure~\ref{fig:observation}, we also present the distribution for high-mass planets. In the case of high-mass planets, the range of model predictions is narrower than that for low-mass planets. Moreover, our model predictions for low-mass planets, in contrast to Jupiter-mass planets, generally exhibit lower equivalent width. This is consistent with the frequent non-detections of helium absorption in sub-Neptunes, which may be attributed either to a low level of EUV flux or to the cooling effects of metals reducing the atmospheric temperature. 
We computed the mean square error of the predictions. Specifically, we measure the mean square of the residuals in log space as:
\begin{equation}
    \mathrm{RMS_{log}} = \sqrt{\sum_{i=1}^{N}(\log(\mathrm{EW_{obs,i}}) - \log(\mathrm{EW_{pred,i}}))^2} /N
\end{equation}
where $\mathrm{EW_{obs,i}},\mathrm{EW_{pred,i}}$ are equivalent width of observations and predictions and $N$ is the total number of observed planets.
The corresponding relative error of our models $\exp(\mathrm{RMS_{log}})$ is $\sim5,7$ for low-mass and high-mass planets, respectively. The errors are similar in both cases. 

\begin{figure}
    \centering
    \includegraphics[width=1.\linewidth]{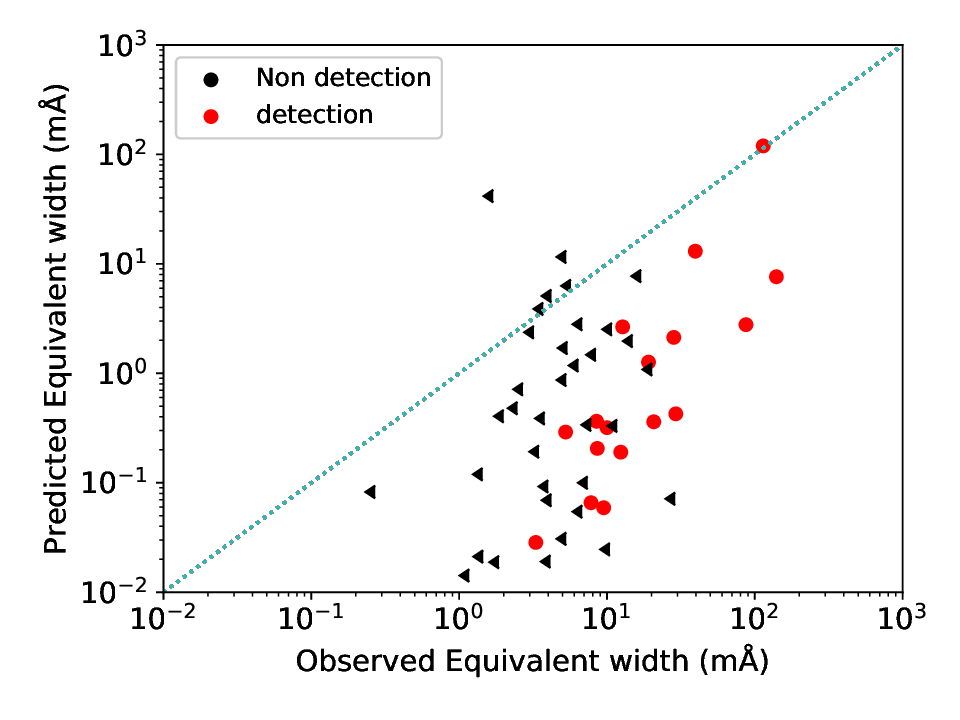}
    \caption{Same as Figure~\ref{fig:observation} but the prediction from Eq.~\ref{eq:Mdot_EW}. }
    \label{fig:observation_prev}
\end{figure}

We also test the prediction from Eq.~\ref{eq:Mdot_EW} in Figure~\ref{fig:observation_prev}. We calculate the mass-loss rate from the EUV flux from MOPYS, assuming the energy-limited mass-loss rate below:
\begin{equation}
    \dot{M} = \epsilon\frac{F_{\rm EUV}\Rp^3 }{G\Mp}
    \label{eq:mdot_elim}
\end{equation}
where $\epsilon=0.1$ and $F_{\rm EUV}$ are the mass-loss efficiency and EUV flux from the host star. Eq.~\ref{eq:mdot_elim} ignores the difference between $\Rp$ and the effective radius of EUV absorption ($R_{\rm EUV}$). However, for the low-mass planets in our sample, we typically find $R_{\rm EUV} \lesssim 1.2\,R_{\rm p}$, so the difference in the resulting mass-loss rates is within a factor of $\sim (R_{\rm EUV}/\Rp)^2 \lesssim 1.4$ (i.e., generally within a factor of $\lesssim 2$).

The prediction from Eq.~\ref{eq:Mdot_EW} tends to underestimate the equivalent width even for planets with detections although the relative error is $\sim6$ and close to our models. This trend is inconsistent with the model assumptions that the planets have pure hydrogen/helium dominated atmosphere because metal cooling process and/or stellar wind confinement reduce the mass-loss rate and the equivalent width in real systems. 
Our model prediction is more consistent not only with the hydrodynamic simulations but also with observations.

The parameter $a=G\Mp/\cs^2$ plays a pivotal role in determining the equivalent width, serving as a key diagnostic of the underlying atmospheric dynamics. Interestingly, our model predicts substantial equivalent width values (exceeding 100 m\AA) for planets such as 55 Cnc e, AU Mic b, and TOI01235 b, despite the absence of detectable helium absorption. This discrepancy may indicate the presence of non-standard processes, such as the impact of a strong stellar wind or the possibility of atmospheres that are not hydrogen-dominated. For example, 55 Cnc e is a ultra-short-period exoplanet and hosts a magma ocean. In such systems, the primordial hydrogen atmosphere quickly escapes and a long equivalent width is expected if they have a hydrogen atmosphere. In the case of TOI 1235 b, the planet can also lose the primordial atmosphere \citep{Krishnamurthy_2023}. 
It is difficult to distinguish these planets from the previous simple relationships in Eq.~\ref{eq:Mdot_EW} and the energy-limited mass-loss rates but our model prediction is useful to distinguish them from helium observations.
In the case of AU Mic b, the stellar wind confinement can modify the outflow structure and reduce the observed signals \citep{McCann_2019,Mitani_2022}.  Our model is also useful to find the peculiar non-detections from observations.

\subsection{Other metal cooling effects}
We compare our analytical model with the results of hydrodynamic simulations. In these simulations, we include only Mg-ion radiative cooling as the metal cooling mechanism.

In real atmospheres, radiative cooling by multiple metal species can further reduce the temperature, as demonstrated in previous studies \citep{Zhang_2023c, Linssen_2024, Kubyshkina_2024, Fossati_2025}. To assess this effect, we compute the metal cooling rate in post-processing using the \texttt{sunbather} module \citep{Linssen_2024}, which is based on the open-source code CLOUDY \citep{Chatzikos_2023} to check the fact that the atmospheric temperature is reduced by metal coolings in real systems.
We note that our post-processing is not used for the self-consistent calculations but we don’t expect the changes in the radiative heating/temperature to affect significantly the mass-loss rate and density.

\begin{figure}
    \centering
    \includegraphics[width=\linewidth]{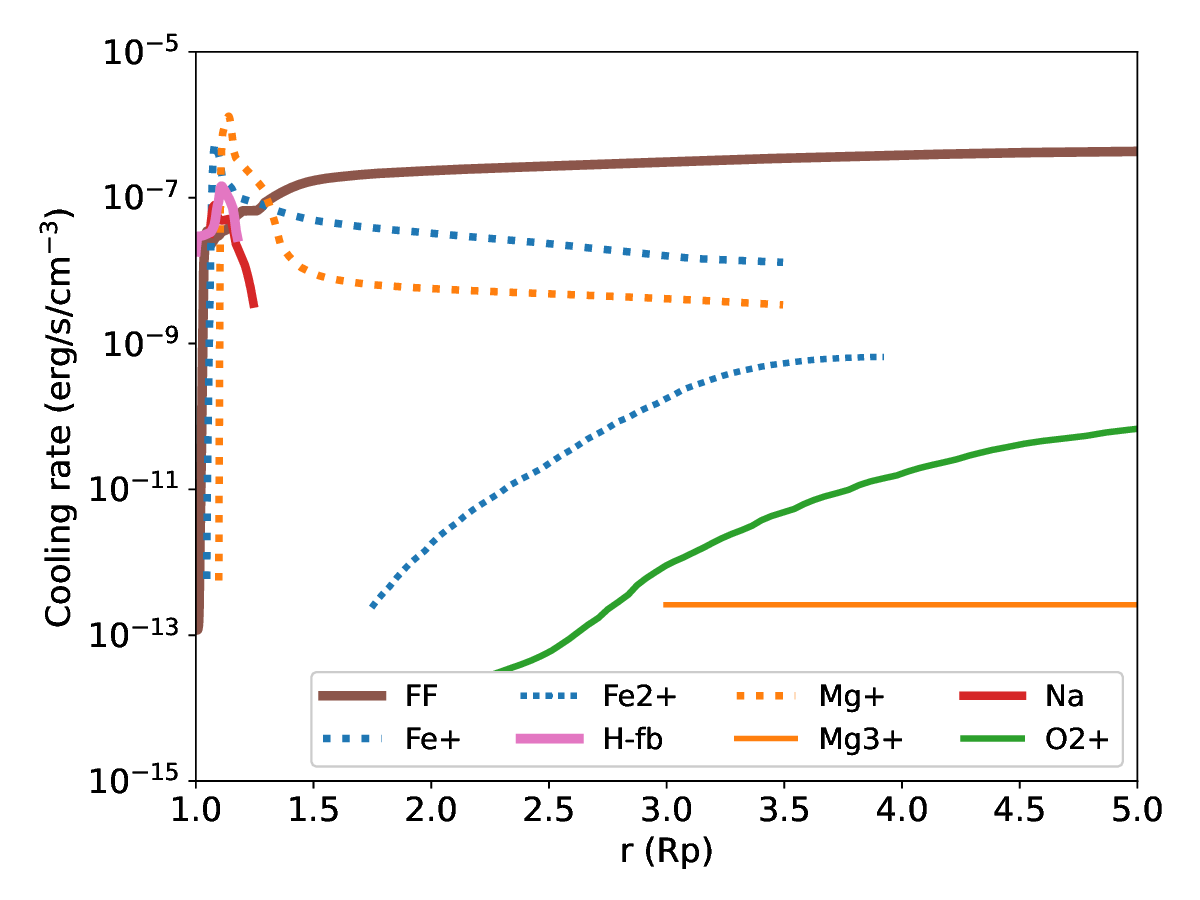}
    \caption{Radial profile of cooling rate of $Z = 10 Z_{\odot}$ case from the post-processing calculation. Only those with a fraction greater than 0.1 are shown but other minor cooling processes not shown in the figure are also calculated. Free-free and free-bound cooling are also shown. }
    \label{fig:cloudy}
\end{figure}

Figure~\ref{fig:cloudy} shows the radial profile of cooling fractions for the $Z = 10 Z_{\odot}$ case. Near the surface, Mg and Fe ions are the dominant cooling sources, while oxygen ions become dominant in the outer regions. Although the identity of the dominant cooling species varies with metallicity, we find that the overall temperature profile remains nearly independent of which metal ion dominates.

\begin{figure}
    \centering
    \includegraphics[width=\linewidth]{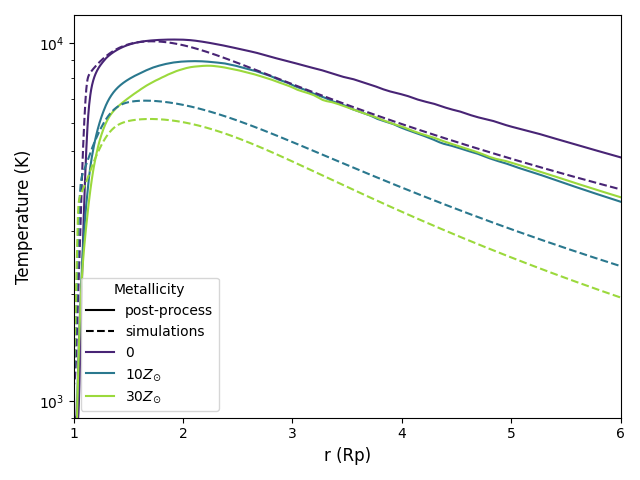}
    \caption{Atmospheric temperature profiles of post-processing calculations with different metallicity $Z=0,10Z_{\odot},30Z_{\odot}$ cases. The solid and dashed lines represent profiles of post-processing calculation and simulations in Figure~\ref{fig:profile_HJ}. }
    \label{fig:cloudy_temp}
\end{figure}

In our post-processing calculations, cooling by metal species other than Mg becomes dominant. Figure~\ref{fig:cloudy_temp} shows the atmospheric temperature profiles for different metallicities. Temperatures in metal-rich planets are systematically higher in post-processing because, in addition to EUV hydrogen photoionization heating, we include other photoheating processes; in particular, heating by oxygen ions contributes roughly 10 \% of the total heating throughout the atmosphere of these planets. The atmospheric temperature in real systems might be higher ($>5000\mathrm{\,K}$) and low mass-loss rate and long equivalent width can be rare.
Despite this, the metallicity dependence of the maximum atmospheric temperature closely mirrors that of the Mg-only cooling case, since radiative cooling becomes inefficient at lower temperatures.

We note that our analytical models assume the isothermal profile of the atmosphere and isotropic helium metastable state fraction and are independent of the dominant cooling process.

\section{Summary}
Helium triplet absorption has been widely used to observe the atmospheric escape of close-in exoplanets.
Understanding how helium absorption depends on the temperature of a planetary atmosphere is important for interpreting observations of different metallicity.
We derived analytical models for helium absorption which can be used for general close-in planets with a hydrogen-dominated atmosphere. Our analytical models only assume the upper part of the atmosphere to be isothermal and we assume an isotropic fraction of metastable helium; the model is generally applicable for close-in planets around low-mass stars with hydrogen-dominated atmospheres.
We also perform hydrodynamics simulations and find that our analytical model predicts a weak dependence of equivalent width on the mass-loss rate in the case of hot Jupiters which is consistent with the simulations.

We also compare our 1D radiation-hydrodynamics model with previous estimates based on the energy-limited mass-loss approach. We find that the latter tends to underestimate the equivalent width. In contrast to previous estimates in \cite{Zhang_2022,Zhang_2023c}, our model provides predictions that are more consistent with the observed values. Our model also predicts longer equivalent widths for ultra-short period planets. If the planets possessed a primordial hydrogen-dominated atmosphere, the strong outflow launched by intense radiation shows deep absorption. The simple estimation of mass-loss rate from the equivalent width overestimates the mass-loss rate.
This comparison underscores the necessity of incorporating metal cooling and its associated thermo-chemical effects to accurately interpret helium triplet observations. We test the different metal cooling species and find that the maximum temperature for Mg-only is similar to that with other metal species because the metal radiative cooling becomes inefficient at lower temperatures. Our analytical model only uses an assumption on the temperature of the upper atmosphere and is therefore independent of the specific dominant cooling process. Our model provides an easy way to estimate the helium triplet fraction of the systems and can also be used to identify planets with helium observations that cannot be explained by simple hydrogen and helium atmospheres. We also find that it is difficult for a simple estimate from the energy-limited mass-loss rate to identify such planets.  In such systems, non-standard processes such as stellar wind compression or the absence of a primordial hydrogen-rich atmosphere may be required to understand the thermo-chemical structures. 

In short, our study presents a comprehensive framework for understanding the interplay between metal species, atmospheric structure, and atmospheric escape processes for close-in exoplanets. By accounting for the influence of metal cooling, we offer a refined interpretation of helium triplet absorption observations. Future work combining detailed observations with advanced modeling will be essential to further unravel the complex dynamics governing exoplanetary atmospheres.

\begin{acknowledgements}
We thank the anonymous referee for constructive comments and suggestions, which helped to improve the quality and clarity of this paper.
HM has been supported by JSPS Overseas Research Fellowship. This work was supported by JSPS KAKENHI Grant Number 25K17432. Numerical computations were in part carried out on Cray XD2000 at the Center for Computational Astrophysics, National Astronomical Observatory of Japan.
RK acknowledges financial support via the Heisenberg Research Grant funded by the Deutsche Forschungsgemeinschaft (DFG, German Research Foundation) under grant no.~KU 2849/9, project no.~445783058. 
\end{acknowledgements}

\bibliographystyle{aa}
\bibliography{references}
\begin{appendix}
\section{The metastable helium fraction}
\begin{figure}[h]
    \centering
    \includegraphics[width=\linewidth]{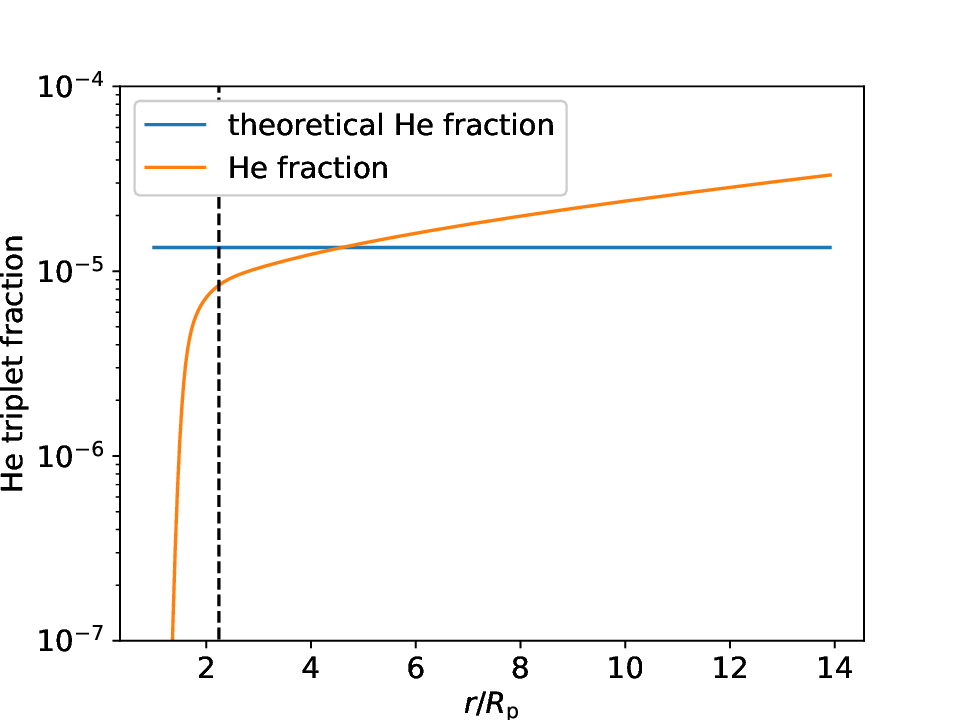}
    \caption{The helium metastable fraction of low-mass planets with $Z=10Z_{\odot}$. The solid lines represent the simulation result (orange) and the constant fraction from the thermospheric balance temperature. The dashed line shows the position of the sonic point.}
    \label{fig:he_frac}
\end{figure}

In our analytical model, we assume a constant metastable helium fraction because we focus on young planetary systems around low-mass stars, where the assumption of a constant $f_3$ with height is valid. Figure~\ref{fig:he_frac} shows the $f_3$ profile from our simulations of low-mass planets. The effect of the altitude dependence of $f_3$ remains within a factor of 2 across almost the entire region of our simulations. 

\section{Planetary parameters and model predictions}\label{app:B}

This appendix summarizes the planetary parameters adopted in the modeling and
the predicted equivalent widths shown in Fig.~\ref{fig:observation}. The observational data are taken from the MOPYS compilation \citep{Orell-Miquel_2024}, and we compute the equivalent width from the published measurements.

\begin{sidewaystable*}[t]
\centering
\caption{Low-mass planet samples from MOPYS \citep{Orell-Miquel_2024}. $EW_{\rm pred}$ represents our model predictions. The symbol “N” denotes that the value of $F_{\rm XUV}$ is not available from MOPYS dataset.}
\scriptsize
\setlength{\tabcolsep}{3pt}
\renewcommand{\arraystretch}{1.1}
\resizebox{\textwidth}{!}{%
\begin{tabular}{lrrrrrrrrrrr}
\hline
Planet & $P$ (day) & $R_{\rm p}$ ($R_\oplus$) & $M_{\rm p}$ ($M_\oplus$) & Age (Gyr) & $M_\star$ ($M_\odot$) & $R_\star$ ($R_\odot$) & $T_{\rm eff}$ (K) & $D_{\rm He}$ (percentile) & $F_{\rm XUV}$ (W m$^{-2}$) & $EW_{\rm He}$ ($\mathrm{m\AA}$) & $EW_{\rm pred} (\mathrm{m\AA})$\\
\hline
55\_Cnc\_e & $0.7365$ & $1.875\,\pm\,0.029$ & $7.99^{+0.32}_{-0.33}$ & $10.2\,\pm\,2.5$ & $0.905\,\pm\,0.015$ & $0.943\,\pm\,0.01$ & $5172\,\pm\,18$ & $<0.025$ & $4.57$ & $<0.27$ & $162$ \\
AU\_Mic\_b & $8.463$ & $4.16\,\pm\,0.18$ & $11.7\,\pm\,5.0$ & $0.022\,\pm\,0.003$ & $0.5\,\pm\,0.03$ & $0.75\,\pm\,0.03$ & $3700\,\pm\,100$ & $<0.34$ & $22.9$ & $<3.7$ & $1.91\times 10^{3}$ \\
GJ436\_b & $2.6441$ & $4.19\,\pm\,0.1$ & $23.14\,\pm\,0.76$ & $6.0^{+4.0}_{-3.0}$ & $0.445\,\pm\,0.044$ & $0.449\,\pm\,0.019$ & $3479\,\pm\,60$ & $<0.41$ & $0.145$ & $<1.45$ & $0.0972$ \\
GJ806\_b & $0.9263$ & $1.331\,\pm\,0.023$ & $1.9\,\pm\,0.17$ & $4.0^{+4.0}_{-3.0}$ & $0.413\,\pm\,0.011$ & $0.4144\,\pm\,0.0038$ & $3600\,\pm\,16$ & $<0.7$ & $1.62$ & $<7.4$ & $0.151$ \\
GJ1214\_b & $1.5804$ & $2.742^{+0.05}_{-0.053}$ & $8.17\,\pm\,0.43$ & $3.0^{+7.0}_{-0.0}$ & $0.178\,\pm\,0.01$ & $0.215\,\pm\,0.008$ & $3250\,\pm\,100$ & $<2.1^{+0.45}_{-0.5}$ & $0.2951$ & $<28.9^{+9.4}_{-8.5}$ & $0.519$ \\
GJ3470\_b & $3.3366$ & $4.04\,\pm\,0.11$ & $11.44\,\pm\,0.64$ & $1.65\,\pm\,1.4$ & $0.476\,\pm\,0.019$ & $0.474\,\pm\,0.014$ & $3725\,\pm\,54$ & $1.5\,\pm\,0.3$ & $1.44$ & $20.72\,\pm\,1.3$ & $18.9$ \\
GJ9827\_b & $1.209$ & $1.529\,\pm\,0.058$ & $4.87\,\pm\,0.37$ & $10.0^{+3.0}_{-5.0}$ & $0.593\,\pm\,0.018$ & $0.579\,\pm\,0.018$ & $4294\,\pm\,52$ & $<0.21$ & $0.72$ & $<1.86$ & $7.3$ \\
GJ9827\_d & $6.2018$ & $1.955\,\pm\,0.075$ & $3.42\,\pm\,0.62$ & $10.0^{+3.0}_{-5.0}$ & $0.593\,\pm\,0.018$ & $0.579\,\pm\,0.018$ & $4294\,\pm\,52$ & $<0.3$ & $0.081$ & $<1.24$ & $0.00172$ \\
HD63433\_b & $7.1079$ & $2.141^{+0.09}_{-0.067}$ & $5.5\,\pm\,2.3$ & $0.414\,\pm\,0.023$ & $0.956\,\pm\,0.022$ & $0.934\,\pm\,0.029$ & $5553\,\pm\,56$ & $<0.34$ & $10.3$ & $<2.0$ & $1.9$ \\
HD63433\_c & $20.5438$ & $2.69^{+0.11}_{-0.09}$ & $15.5^{+3.9}_{-3.8}$ & $0.414\,\pm\,0.023$ & $0.956\,\pm\,0.022$ & $0.934\,\pm\,0.029$ & $5553\,\pm\,56$ & $<0.4$ & $2.5$ & $<4.2$ & $11.3$ \\
HD73583\_b & $6.398$ & $2.79\,\pm\,0.1$ & $10.2^{+3.4}_{-3.1}$ & $0.49\,\pm\,0.19$ & $0.71\,\pm\,0.02$ & $0.66\,\pm\,0.02$ & $4511\,\pm\,110$ & $0.72\,\pm\,0.08$ & $3.1$ & $8.6\,\pm\,0.6$ & $46.3$ \\
HD73583\_c & $18.8797$ & $2.39\,\pm\,0.1$ & $9.7^{+1.8}_{-1.7}$ & $0.49\,\pm\,0.19$ & $0.71\,\pm\,0.02$ & $0.66\,\pm\,0.02$ & $4511\,\pm\,110$ & $<0.5$ & $0.7$ & $<5.3$ & $3.26$ \\
HD97658\_b & $9.4893$ & $2.247^{+0.098}_{-0.095}$ & $7.55^{+0.83}_{-0.79}$ & $6.0\,\pm\,1.0$ & $0.77\,\pm\,0.05$ & $0.741\,\pm\,0.024$ & $5170^{+50}_{-0}$ & $<0.21$ & $0.128$ & $<0.9$ & $0.594$ \\
HD235088\_b & $7.4341$ & $2.045\,\pm\,0.075$ & $7.0\,\pm\,2.0$ & $0.65^{+0.15}_{-0.05}$ & $0.843\,\pm\,0.05$ & $0.789\,\pm\,0.021$ & $5037\,\pm\,14$ & $0.91^{+0.11}_{-0.1}$ & $1.854$ & $9.5^{+1.1}_{-1.0}$ & $33$ \\
HAT-P-11\_b & $4.8878$ & $4.36\,\pm\,0.06$ & $27.7\,\pm\,3.1$ & $6.5^{+5.9}_{-4.1}$ & $0.802\,\pm\,0.028$ & $0.683\,\pm\,0.009$ & $4780\,\pm\,50$ & $1.08\,\pm\,0.05$ & $2.109$ & $12.4\,\pm\,2.4$ & $4.96$ \\
LTT9779\_b & $0.792$ & $4.72\,\pm\,0.23$ & $28.32^{+0.78}_{-0.81}$ & $2^{+1.3}_{-0.9}$ & $1.02\,\pm\,0.03$ & $0.949\,\pm\,0.006$ & $5443\,\pm\,14$ & $<0.2$ & $4.8$ & $<3.79$ & $60.7$ \\
TOI-1136\_d & $12.5194$ & $4.626^{+0.076}_{-0.072}$ & $8.0^{+2.4}_{-1.9}$ & $0.7\,\pm\,0.15$ & $1.022\,\pm\,0.027$ & $0.968\,\pm\,0.036$ & $5770\,\pm\,50$ & $<0.5$ & $N$ & $<5.4$ & $6.97$ \\
TOI-1235\_b & $3.4447$ & $1.694^{+0.08}_{-0.077}$ & $5.9^{+0.62}_{-0.61}$ & $5.0^{+5.0}_{-4.4}$ & $0.63\,\pm\,0.024$ & $0.619\,\pm\,0.019$ & $3997\,\pm\,51$ & $<0.09$ & $N$ & $<1.44$ & $76.3$ \\
TOI-1683\_b & $3.0575$ & $2.3\,\pm\,0.3$ & $8.0\,\pm\,3.0$ & $2.0^{+1.3}_{-0.9}$ & $0.69\,\pm\,0.09$ & $0.636\,\pm\,0.03$ & $4539\,\pm\,100$ & $0.84\,\pm\,0.17$ & $7.4$ & $8.5\,\pm\,1.6$ & $197$ \\
TOI-1728\_b & $3.4914$ & $4.62\,\pm\,0.09$ & $26.8^{+5.4}_{-5.1}$ & $7.1\,\pm\,4.6$ & $0.646\,\pm\,0.023$ & $0.6243\,\pm\,0.01$ & $3980^{+31}_{-32}$ & $<1.1$ & $N$ & $<11.7$ & $9.45$ \\
TOI-1807\_b & $0.5494$ & $1.37\,\pm\,0.09$ & $2.57\,\pm\,0.5$ & $0.3\,\pm\,0.08$ & $0.75\,\pm\,0.025$ & $0.68\,\pm\,0.015$ & $4757^{+51}_{-50}$ & $<0.38$ & $3.05$ & $<4.0$ & $2.28$ \\
TOI-2018\_b & $7.4356$ & $2.268\,\pm\,0.069$ & $9.2\,\pm\,2.1$ & $2.4^{+0.2}_{-0.8}$ & $0.57\,\pm\,0.02$ & $0.62\,\pm\,0.01$ & $4174^{+34}_{-42}$ & $1.02^{+0.19}_{-0.22}$ & $1.56$ & $7.8\,\pm\,1.5$ & $10.6$ \\
TOI-2048\_b & $13.7905$ & $2.6\,\pm\,0.2$ & $9.0\,\pm\,3.0$ & $0.3\,\pm\,0.05$ & $0.83\,\pm\,0.03$ & $0.79\,\pm\,0.04$ & $5185\,\pm\,60$ & $<1.0$ & $N$ & $<10.2$ & $933$ \\
TOI-2076\_b & $10.3557$ & $2.52\,\pm\,0.056$ & $9.0\,\pm\,3.0$ & $0.34\,\pm\,0.08$ & $0.824\,\pm\,0.036$ & $0.77\,\pm\,0.006$ & $5200\,\pm\,70$ & $1.01\,\pm\,0.05$ & $6.7$ & $10.0\,\pm\,0.7$ & $195$ \\
TOI-2134\_b & $9.2292$ & $2.69\,\pm\,0.16$ & $9.13^{+0.78}_{-0.76}$ & $3.8^{+5.5}_{-2.7}$ & $0.744\,\pm\,0.027$ & $0.709\,\pm\,0.017$ & $4580\,\pm\,54$ & $0.38\,\pm\,0.05$ & $0.46$ & $3.3\,\pm\,0.3$ & $3.51$ \\
TOI-2136\_b & $7.8519$ & $2.2\,\pm\,0.07$ & $4.7^{+3.1}_{-2.6}$ & $4.6\,\pm\,1.0$ & $0.3272\,\pm\,0.0082$ & $0.344\,\pm\,0.0099$ & $3373\,\pm\,108$ & $<1.44$ & $N$ & $<7.8$ & $0.404$ \\
TRAPPIST-1\_b & $1.5109$ & $1.086\,\pm\,0.035$ & $0.85\,\pm\,0.72$ & $7.6\,\pm\,2.2$ & $0.0802\,\pm\,0.0073$ & $0.117\,\pm\,0.0036$ & $2550\,\pm\,50$ & $<0.33$ & $0.7244$ & $<3.467$ & $0.0578$ \\
TRAPPIST-1\_e & $6.0996$ & $0.918\,\pm\,0.039$ & $0.62\,\pm\,0.58$ & $7.6\,\pm\,2.2$ & $0.0802\,\pm\,0.0073$ & $0.117\,\pm\,0.0036$ & $2550\,\pm\,50$ & $<1.07$ & $0.112$ & $<10.458$ & $0.00185$ \\
TRAPPIST-1\_f & $9.2067$ & $1.045\,\pm\,0.038$ & $0.68\,\pm\,0.18$ & $7.6\,\pm\,2.2$ & $0.0802\,\pm\,0.0073$ & $0.117\,\pm\,0.0036$ & $2550\,\pm\,50$ & $<0.38$ & $0.0645$ & $<4.143$ & $0.00103$ \\
K2-25\_b & $3.4846$ & $3.43\,\pm\,0.12$ & $28.5^{+8.5}_{-8.3}$ & $0.725\,\pm\,0.075$ & $0.294\,\pm\,0.021$ & $0.295\,\pm\,0.02$ & $3180\,\pm\,60$ & $<1.7$ & $N$ & $<17.0$ & $99.3$ \\
K2-77\_b & $8.1998$ & $2.3\,\pm\,0.16$ & $9.0^{+600.0}_{-1.0}$ & $0.12^{+0.78}_{-0.02}$ & $0.8\,\pm\,0.12$ & $0.76\,\pm\,0.03$ & $4970\,\pm\,45$ & $<2.7$ & $13.45$ & $<28.0$ & $406$ \\
K2-100\_b & $1.6739$ & $3.88\,\pm\,0.16$ & $21.8\,\pm\,6.2$ & $0.7\,\pm\,0.1$ & $1.15\,\pm\,0.05$ & $1.24\,\pm\,0.05$ & $5945\,\pm\,110$ & $<1.3$ & $141.253$ & $<5.7$ & $1.21\times 10^{4}$ \\
K2-105\_b & $8.267$ & $3.59^{+0.11}_{-0.07}$ & $30.0\,\pm\,19.0$ & $5.0^{+8.0}_{-4.4}$ & $1.05\,\pm\,0.02$ & $0.97\,\pm\,0.01$ & $5636^{+49}_{-52}$ & $<2.33$ & $14.69$ & $<24.7$ & $28.5$ \\
K2-136\_c & $17.307$ & $3.0\,\pm\,0.13$ & $18.1^{+1.8}_{-1.9}$ & $0.65\,\pm\,0.07$ & $0.742\,\pm\,0.02$ & $0.677\,\pm\,0.027$ & $4500\,\pm\,50$ & $<2.3$ & $0.5888$ & $<25.0$ & $0.633$ \\
Kepler-25\_c & $12.7204$ & $5.217^{+0.07}_{-0.065}$ & $15.2^{+1.3}_{-1.6}$ & $2.75\,\pm\,0.3$ & $1.26\,\pm\,0.03$ & $1.34\,\pm\,0.01$ & $6354\,\pm\,27$ & $<1.86$ & $1.019$ & $<19.8$ & $52.1$ \\
Kepler-68\_b & $5.3988$ & $2.31^{+0.06}_{-0.09}$ & $8.3^{+2.2}_{-2.4}$ & $6.3\,\pm\,1.7$ & $1.079\,\pm\,0.051$ & $1.243\,\pm\,0.019$ & $5793\,\pm\,74$ & $<0.72$ & $1.176$ & $<7.6$ & $37.9$ \\
WASP-47\_d & $9.0305$ & $3.567\,\pm\,0.045$ & $14.2\,\pm\,1.3$ & $6.5^{+2.6}_{-1.2}$ & $1.04\,\pm\,0.031$ & $1.137\,\pm\,0.013$ & $5552\,\pm\,75$ & $<3.29$ & $0.577$ & $<34.9$ & $7.46$ \\
\hline
\end{tabular}}

\label{tab:planet_ew_low}
\end{sidewaystable*}

\begin{sidewaystable*}[t]
\centering
\caption{Same as table~\ref{tab:planet_ew_low} but for high-mass planets.}
\scriptsize
\setlength{\tabcolsep}{3pt}
\renewcommand{\arraystretch}{1.1}
\resizebox{\textwidth}{!}{%
\begin{tabular}{lrrrrrrrrrrr}
\hline
Planet & $P$ (day) & $R_{\rm p}$ ($R_\oplus$) & $M_{\rm p}$ ($M_\oplus$) & Age (Gyr) & $M_\star$ ($M_\odot$) & $R_\star$ ($R_\odot$) & $T_{\rm eff}$ (K) & $D_{\rm He}$ (percentile) & $F_{\rm XUV}$ (W m$^{-2}$) & $EW_{\rm He}$ ($\mathrm{m\AA}$) & $EW_{\rm pred} (\mathrm{m\AA})$ \\
\hline
HD89345\_b & $11.8144$ & $6.86\,\pm\,0.14$ & $35.6\,\pm\,3.2$ & $9.4^{+0.4}_{-1.3}$ & $1.12\,\pm\,0.04$ & $1.657\,\pm\,0.02$ & $5499\,\pm\,73$ & $<0.7$ & $0.244$ & $<7.4$ & $1.78$ \\
HD189733\_b & $2.2186$ & $12.76\,\pm\,0.3$ & $357.0\,\pm\,14.0$ & $6.8\,\pm\,5.2$ & $0.806\,\pm\,0.048$ & $0.756\,\pm\,0.018$ & $5040\,\pm\,50$ & $0.75\,\pm\,0.03$ & $16.75$ & $12.76\,\pm\,0.4$ & $3.67$ \\
HD209458\_b & $3.5247$ & $15.23^{+0.16}_{-0.21}$ & $217.7^{+4.8}_{-4.4}$ & $4.0\,\pm\,2.0$ & $1.119\,\pm\,0.033$ & $1.155\,\pm\,0.015$ & $6065\,\pm\,50$ & $0.91\,\pm\,0.1$ & $1.004$ & $5.252\,\pm\,0.5$ & $0.228$ \\
HAT-P-3\_b & $2.8997$ & $10.2\,\pm\,0.4$ & $189.1\,\pm\,7.6$ & $2.9^{+4.9}_{-2.7}$ & $0.925\,\pm\,0.046$ & $0.85\,\pm\,0.021$ & $5190\,\pm\,80$ & $<1.9$ & $7.968$ & $<20.2$ & $0.86$ \\
HAT-P-18\_b & $5.508$ & $11.15\,\pm\,0.58$ & $62.6\,\pm\,4.1$ & $12.4^{+1.4}_{-6.4}$ & $0.77\,\pm\,0.031$ & $0.749\,\pm\,0.037$ & $4803\,\pm\,80$ & $0.7\,\pm\,0.16$ & $0.7$ & $29.21\,\pm\,1.0$ & $3.27$ \\
HAT-P-32\_b & $2.15$ & $20.05\,\pm\,0.28$ & $185.9\,\pm\,9.9$ & $2.7\,\pm\,0.8$ & $1.16\,\pm\,0.041$ & $1.219\,\pm\,0.016$ & $6269\,\pm\,64$ & $5.3\,\pm\,0.1$ & $163$ & $114.0\,\pm\,4.0$ & $3.23\times 10^{3}$ \\
HAT-P-33\_b & $3.4745$ & $18.9\,\pm\,0.5$ & $229.0^{+41.0}_{-38.0}$ & $2.3\,\pm\,0.3$ & $1.42\,\pm\,0.15$ & $1.91\,\pm\,0.26$ & $6460^{+300}_{-290}$ & $<1.4$ & $6.195$ & $<14.9$ & $8.81$ \\
HAT-P-49\_b & $2.6916$ & $15.84^{+1.4}_{-0.86}$ & $550.0\,\pm\,65.0$ & $1.5\,\pm\,0.2$ & $1.543\,\pm\,0.051$ & $1.833\,\pm\,0.138$ & $6820\,\pm\,52$ & $<0.6$ & $14.51$ & $<6.4$ & $1.3$ \\
HAT-P-57\_b & $2.4653$ & $15.84\,\pm\,0.61$ & $588.0^{+590.0}_{-0.0}$ & $1.0^{+0.67}_{-0.51}$ & $1.47\,\pm\,0.12$ & $1.5\,\pm\,0.05$ & $7500\,\pm\,250$ & $<1.0$ & $N$ & $<10.6$ & $4.69$ \\
HAT-P-67\_b & $4.8101$ & $23.37^{+1.1}_{-0.8}$ & $108.0^{+79.0}_{-60.0}$ & $1.24^{+0.24}_{-0.22}$ & $1.642\,\pm\,0.1$ & $2.65\,\pm\,0.12$ & $6406^{+65}_{-61}$ & $10.0\,\pm\,0.1$ & $N$ & $140.0\,\pm\,10.0$ & $2.03\times 10^{3}$ \\
HAT-P-70\_b & $2.7443$ & $21.0^{+1.7}_{-1.1}$ & $<2155$ & $0.60^{+0.38}_{-0.2}$ & $1.890\,\pm\,0.013$ & $1.86\,\pm\,0.12$ & $8450\,\pm\,540$ & $N$ & $N$ & $N$ & $4.45$ \\
MASCARA-2\_b & $3.4741$ & $20.51\,\pm\,0.78$ & $1075.0^{+1100.0}_{-0.0}$ & $0.2^{+0.1}_{-0.05}$ & $1.89\,\pm\,0.06$ & $1.6\,\pm\,0.06$ & $8980^{+90}_{-130}$ & $<0.5$ & $N$ & $<5.3$ & $36.4$ \\
TOI-1268\_b & $8.1577$ & $9.1\,\pm\,0.6$ & $96.0\,\pm\,13.0$ & $0.245\,\pm\,0.14$ & $0.96\,\pm\,0.04$ & $0.92\,\pm\,0.06$ & $5300\,\pm\,100$ & $1.97^{+0.16}_{-0.15}$ & $7.2$ & $19.1^{+1.9}_{-1.8}$ & $5.02$ \\
TOI-1431\_b & $2.6502$ & $16.7\,\pm\,0.56$ & $992.0\,\pm\,57.0$ & $0.29^{+0.32}_{-0.19}$ & $1.895\,\pm\,0.1$ & $1.923\,\pm\,0.068$ & $7690^{+400}_{-250}$ & $<0.4$ & $N$ & $<4.2$ & $10.5$ \\
TOI-2046\_b & $1.4972$ & $16.1\,\pm\,1.2$ & $731.0\,\pm\,89.0$ & $0.4^{+0.22}_{-0.3}$ & $1.13\,\pm\,0.19$ & $1.21\,\pm\,0.07$ & $6200\,\pm\,100$ & $<2.9$ & $22.45$ & $<30.5$ & $3.91$ \\
TOI-3757\_b & $3.4388$ & $12.0^{+0.4}_{-0.5}$ & $85.3^{+8.8}_{-8.7}$ & $7.1\,\pm\,4.5$ & $0.64\,\pm\,0.02$ & $0.62\,\pm\,0.01$ & $3913\,\pm\,56$ & $<6.9$ & $N$ & $<73.0$ & $11.2$ \\
NGTS-5\_b & $3.357$ & $12.73\,\pm\,0.26$ & $73.0\,\pm\,12.0$ & $5.0^{+8.0}_{-3.5}$ & $0.661\,\pm\,0.065$ & $0.739\,\pm\,0.014$ & $4987\,\pm\,41$ & $<1.02^{+0.48}_{-0.46}$ & $3.2$ & $<10.8^{+5.1}_{-4.9}$ & $34.9$ \\
KELT-9\_b & $1.4811$ & $21.701\,\pm\,0.053$ & $920.0\,\pm\,110.0$ & $0.45^{+0.14}_{-0.13}$ & $2.32\,\pm\,0.16$ & $2.418\,\pm\,0.058$ & $9600\,\pm\,400$ & $<0.33$ & $0.15$ & $<1.17$ & $0.00771$ \\
V1298Tau\_c & $8.2489$ & $5.2\,\pm\,0.39$ & $76.0^{+76.0}_{-0.0}$ & $0.023\,\pm\,0.004$ & $1.17\,\pm\,0.06$ & $1.278\,\pm\,0.07$ & $5050\,\pm\,100$ & $<3.75$ & $151.356$ & $<95.8$ & $18.1$ \\
V1298Tau\_b & $24.1399$ & $9.77\,\pm\,0.65$ & $203.0\,\pm\,60.0$ & $0.023\,\pm\,0.004$ & $1.17\,\pm\,0.06$ & $1.278\,\pm\,0.07$ & $5050\,\pm\,100$ & $<1.7$ & $87.09$ & $<19.0$ & $6.07$ \\
WASP-11\_b & $3.7225$ & $11.1\,\pm\,0.25$ & $156.0\,\pm\,8.0$ & $7.6^{+6.0}_{-3.5}$ & $0.81\,\pm\,0.04$ & $0.772\,\pm\,0.015$ & $4900\,\pm\,65$ & $<1.56$ & $1.9$ & $<16.6$ & $0.331$ \\
WASP-12\_b & $1.0914$ & $21.71\,\pm\,0.63$ & $466.0\,\pm\,25.0$ & $2.0^{+0.8}_{-1.0}$ & $1.434\,\pm\,0.11$ & $1.657\,\pm\,0.046$ & $6300^{+200}_{-100}$ & $<0.5$ & $3.18$ & $<5.3$ & $1.46$ \\
WASP-39\_b & $4.0553$ & $14.34\,\pm\,0.45$ & $89.0\,\pm\,10.0$ & $8.5^{+4.0}_{-3.4}$ & $0.913\,\pm\,0.047$ & $0.939\,\pm\,0.022$ & $5485\,\pm\,50$ & $<2.04$ & $1.2$ & $<21.7$ & $9.09$ \\
WASP-48\_b & $2.1436$ & $18.7\,\pm\,1.1$ & $328.0^{+15.0}_{-14.0}$ & $7.9^{+2.0}_{-1.6}$ & $1.19\,\pm\,0.05$ & $1.75\,\pm\,0.09$ & $6000\,\pm\,150$ & $<0.25\pm0.21$ & $N$ & $<2.7\pm2.2$ & $0.799$ \\
WASP-52\_b & $1.7498$ & $14.04\,\pm\,0.3$ & $137.9\,\pm\,7.6$ & $0.4^{+0.3}_{-0.2}$ & $0.804\,\pm\,0.05$ & $0.786\,\pm\,0.016$ & $5000\,\pm\,100$ & $3.44\,\pm\,0.31$ & $24.8$ & $39.583\,\pm\,1.4$ & $128$ \\
WASP-69\_b & $3.8681$ & $11.85\,\pm\,0.53$ & $82.6\,\pm\,5.4$ & $2.0\,\pm\,0.5$ & $0.826\,\pm\,0.029$ & $0.813\,\pm\,0.028$ & $4700\,\pm\,50$ & $3.59\,\pm\,0.19$ & $4.17$ & $28.31\,\pm\,0.9$ & $24.7$ \\
WASP-76\_b & $1.8099$ & $20.78^{+0.86}_{-0.85}$ & $284.1^{+4.4}_{-4.1}$ & $5.3^{+6.1}_{-2.9}$ & $1.458\,\pm\,0.021$ & $1.756\,\pm\,0.071$ & $6329\,\pm\,25$ & $<0.88$ & $112.2018$ & $<1.7$ & $351$ \\
WASP-77\_b & $1.36$ & $13.79^{+0.35}_{-0.33}$ & $530.0^{+22.0}_{-20.0}$ & $6.2^{+4.0}_{-3.5}$ & $0.903\,\pm\,0.06$ & $0.91\,\pm\,0.025$ & $5617\,\pm\,72$ & $<0.8$ & $13.182$ & $<8.4$ & $2.03$ \\
WASP-80\_b & $3.0679$ & $11.2^{+0.35}_{-0.34}$ & $171.0\,\pm\,11.0$ & $0.1^{+0.1}_{-0.0}$ & $0.577\,\pm\,0.05$ & $0.586\,\pm\,0.018$ & $4143^{+92}_{-94}$ & $<0.85$ & $1.6595$ & $<2.48$ & $0.318$ \\
WASP-107\_b & $5.7215$ & $10.54\,\pm\,0.22$ & $35.0\,\pm\,3.2$ & $8.3\,\pm\,4.3$ & $0.69\,\pm\,0.05$ & $0.66\,\pm\,0.02$ & $4430\,\pm\,120$ & $7.26\,\pm\,0.24$ & $2.664$ & $87.152\,\pm\,7.6$ & $75$ \\
WASP-127\_b & $4.1781$ & $15.36\,\pm\,0.45$ & $57.2\,\pm\,6.4$ & $11.41\,\pm\,1.8$ & $1.08\,\pm\,0.03$ & $1.39\,\pm\,0.03$ & $5620\,\pm\,85$ & $<0.48$ & $0.058$ & $<6.8$ & $0.812$ \\
WASP-177\_b & $3.0717$ & $17.7^{+7.4}_{-4.0}$ & $161.0\,\pm\,12.0$ & $9.7\,\pm\,3.9$ & $0.876\,\pm\,0.038$ & $0.885\,\pm\,0.046$ & $5017\,\pm\,70$ & $<1.28^{+0.3}_{-0.29}$ & $3.5$ & $<6.8\pm1.6$ & $14.5$ \\
WASP-189\_b & $2.724$ & $18.15\,\pm\,0.24$ & $632.0^{+60.0}_{-44.0}$ & $0.75\,\pm\,0.13$ & $2.03\,\pm\,0.066$ & $2.36\,\pm\,0.03$ & $8000\,\pm\,80$ & $<0.3$ & $N$ & $<3.2$ & $4.15$ \\
\hline
\end{tabular}}
\caption{}
\label{tab:planet_ew_high}
\end{sidewaystable*}

\end{appendix}

\end{document}